\documentclass[journal,comsoc]{IEEEtran}

\usepackage{cite}

\usepackage{amsmath}
\usepackage{amsfonts}
\usepackage{amssymb}
\usepackage{cases}
\usepackage{graphicx}
\usepackage{subcaption}
\usepackage{scalerel}
\usepackage{tikz}
\usetikzlibrary{svg.path}
\usepackage{amsmath}
\usepackage{bbm}
\usepackage{booktabs}
\usepackage{graphicx}
\usepackage{mathptmx}
\usepackage[mathscr]{euscript}
\usepackage{amsmath}
\usepackage{mathtools,amssymb,lipsum}
\usepackage{geometry}
\usepackage{diagbox} % add this in your preamble
\usepackage{cuted}
\usepackage{lipsum}
\usepackage{hhline}
\usepackage{hyperref}
\usepackage{caption}
\usepackage{subcaption}
\usepackage{float}
\usepackage[T1]{fontenc}
\usepackage{amssymb}
\usepackage{mathtools}
\usepackage{fixltx2e}
\usepackage{cite}
\usepackage{lettrine}
\usepackage{tabularx,booktabs}
\usepackage[english]{babel}
\usepackage{algorithm}
\usepackage[noend]{algpseudocode}
\usepackage{siunitx} % For proper typesetting of units
\usepackage{scalerel}
\newcolumntype{C}{>{\centering\arraybackslash}X} % centered version of "X" type
\usepackage{epstopdf}
\usepackage{multicol}
\usepackage{multirow}
\usepackage{amssymb}
\usepackage{mathtools}
\usepackage{cuted}
\makeindex

\usetikzlibrary{shapes,arrows.meta,arrows}

\newcolumntype{P}[1]{>{\centering\arraybackslash}p{#1}}
\definecolor{light_pink}{rgb}{255,153,153}

\begin{document}
% \IEEEoverridecommandlockouts\IEEEpubid{\makebox[\columnwidth]{ 979-8-3503-1090-0/23/\$31.00~\copyright~2023 IEEE \hfill} \hspace{\columnsep}\makebox[\columnwidth]{ }}

\renewcommand{\baselinestretch}{1}
%the previously submitted version had 0.88
%\renewcommand{\baselinestretch}{0.88}
% \title{Energy Efficient Band Switching in Multiband Heterogeneous Networks using LSTM}
%\title{Deep Learning aided Energy Efficient Band Assignment in Multiband Heterogeneous Networks}
%\title{Channel Forecasting aided Energy Efficient Band Assignment in Multiband Heterogeneous Networks}
\title{Forecasting Aided Energy Aware Band \\ Assignment in Multiband  Networks}
%\title{Rate Forecaster based Energy Aware Band Assignment in Multiband  Networks}
%\title{Channel Prediction aided Energy Efficient Band Assignment in Multiband Heterogeneous Networks}
%\title{Channel Forecasting aided Energy Efficient Band Assignment in Sub-6/mmWave Heterogeneous Networks}
%
% \author{\IEEEauthorblockN{Brijesh Soni\IEEEauthorrefmark{1}, Siddhartan Govindasamy\IEEEauthorrefmark{2},  Dhaval K. Patel\IEEEauthorrefmark{3}}
% \IEEEauthorblockA{Department of Engineering, Boston College, USA\IEEEauthorrefmark{1}\IEEEauthorrefmark{2}\\
% School of Engineering and Applied Science, Ahmedabad University, India\IEEEauthorrefmark{3}\\
% Email: \IEEEauthorrefmark{1}\IEEEauthorrefmark{2}\{sonib, siddhartan.govindasamy\}@bc.edu,  \IEEEauthorrefmark{3}dhaval.patel@ahduni.edu.in}
% \vspace{-4.5mm}
% }

   %\IEEEauthorrefmark{4}sunsm@i2r.a-star.edu.sg,
%  		\IEEEauthorrefmark{5}ycchang@utar.edu.my,
%  		\IEEEauthorrefmark{6}Joanne.Lim@monash.edu}}

\author{Brijesh Soni,~\IEEEmembership{Member,~IEEE,}
        Siddhartan Govindasamy,~\IEEEmembership{Member,~IEEE,}
        Dhaval K. Patel,~\IEEEmembership{Senior Member,~IEEE}
        % <-this % stops a space
%\thanks{Manuscript received on January 06, 2021. Authors would like to thank financial support received from ASEAN under the DST Thematic Partnerships 2017-18 (grant ref. IMRC/AISTDF/R\&D/P-09/2017).} 
\thanks{The results in parts were presented at the IEEE GLOBECOM conference \cite{our_globecom23_paper}, held in December 2023 at Kualalumpur, Malaysia.}
\thanks{Brijesh Soni is with the Department of Computer Science and Engineering, The Ohio State University, OH, USA. (email: {soni.152@osu.edu})}
\thanks{Siddhartan Govindasamy is with the Department of Engineering, Boston College, MA, USA. (email: {siddhartan.govindasamy@bc.edu})}
\thanks{Dhaval K. Patel is with the School of Engineering and Applied Science, Ahmedabad University, India. (email: {dhaval.patel@ahduni.edu.in})}
}
\maketitle

	
\begin{abstract}
%The Sub-6 GHz and higher frequency bands like mmWave, and sub-THz in multiband systems complement each other. 
%Most of the existing works have considered the rate  of user equipment (UE) as a key criterion for the band assignment policy. 
The high frequency communication bands (mmWave and sub-THz) promise tremendous data rates. However, they have very high power consumption which is particularly significant for battery-powered user-equipment (UE), and are prone to blockage. In this context, we design an energy-aware  band-assignment system which reduces power consumption while aiming to achieve a target sum rate of $M$ bits/sec in $T$ time-slots. We do this by using 1) Rate forecaster(s); 2) Channel forecaster(s) which forecast either the data rate or the channel for $T$ subsequent time slots, utilizing either a stacked Long-short-term memory (LSTM) or transformer architecture. These forecasts are used to select frequency bands using an iterative algorithm. The proposed approach is validated on the publicly available  `DeepMIMO', and `NYUSIM' datasets for both outdoor and indoor scenarios, and using a multiband empirical system. Moreover, we also propose a simple blockage segment generation algorithm such that the channel realizations inherently capture the effects of blockage. We find that the rate-forecaster-based approach outperforms the channel forecaster. Further, our approach consumes $\sim 300$ mW lower power compared to a greedy band assignment at a $1.5$ Gb/s target rate for outdoor scenarios, and up to $600$ mW for indoor scenarios at $2.5$ Gb/s target rate, indicating that this is a promising method to reduce UE power consumption in multiband systems.

%RMSE of predicted rate using timeseries split approach is much less than the conventional approach. \textcolor{black}{Moreover, with our proposed scheme, the median energy efficiency is 56.30\% higher as compared to the case when energy usage is not considered in the bandswitching decision, while the median rate with our proposed scheme reduces only by 13.81\%.}
%Moreover, the proposed scheme has an average 13.34\% higher energy efficiency than the existing band assignment policies. %\textcolor{black}{Additionally, our system performs rate prediction based on prior rates and UE position, resulting in an root-mean-square error that is XX\% of the mean rate, which useful for practical systems.}
\end{abstract}


% \fi
\IEEEpeerreviewmaketitle

\section{Introduction}
\label{sec:Introduction}
%\subsection{New introduction}
%Due to the evolutionary nature of wireless technology and services, the demand for wireless radio frequency (RF) spectrum in the Sub-6 GHz band has continually increased since its inception. 
%Spectrum sharing, one of the potential solutions to alleviate the spectrum crunch, has received much attention in the regulatory, industrial and academic communities. 
%Can't miss to add these two ref. regarding survey on multiband \cite{2024TCOM_multiband_survey, 2023_comm_magazine_multiband}.

\IEEEPARstart {M}{ulti-band} \textcolor{black}{wireless networks are emerging as a key enabler for next-generation communication systems, offering the flexibility to exploit spectrum resources across Sub-6 GHz, millimeter (mmWave), and Terahertz (THz) bands to meet diverse service requirements and use cases \cite{2024TCOM_multiband_survey, 2023_comm_magazine_multiband}.} 3GPP has considered significant available bandwidth at mmWave frequencies in its fifth generation new radio (5G-NR) standard, and is potentially also considering  Sub-THz bands (>100 GHz) which would help achieve beyond5G/6G objectives \cite{mmWave_and_THz_for6G_arxiv2021}.
%One of the promising solutions to alleviate the spectrum crunch is to utilize the larger bandwidth available at the millimeter-wave (mmWave) band (and even at Terahertz (THz) band in the near future) \cite{mmWave_and_THz_for6G_arxiv2021}. %\textcolor{black}{(802.15 THz group)} 
Although the mmWave and THz bands provide significant bandwidth and promise tremendously high data rates (ranging from multi-Gb/s up to Tb/s), propagation at these bands is limited by severe path loss,
%higher attenuation, 
and is susceptible to blockages. These factors limit the quality of service and user experience \cite{THZ_last_piece_of_RF_OJCS}. 

Due to the complementary characteristics of wireless signal propagation in Sub-6 GHz, mmWave bands \textcolor{black}{and THz bands}, 3GPP standards have also evolved to support multiband networks \cite{integrated_mmWave_sub6_roadmap_MDebbah_Comm_magazine2019}, leading to the band assignment problem. The \textit{key idea} regarding band switching in multiband heterogeneous networks is that depending on the use case scenario, the user equipment (UE) requests its serving base station (BS) to switch the band, e.g., from Sub-6 GHz to mmWave or vice-versa. 
%Based on certain policies, the BS decides whether to assign the alternate band or not. The signalling for band switching procedure, also referred to as a legacy approach was proposed in \cite{3gpp_36_331}. 
\subsection{Current state of the art and Motivation}
\textcolor{black}{There are several works in the literature that have investigated the band assignment problem in multiband heterogeneous systems. For instance,
%e.g.  \cite{Beyond5G_MAC_HWN_ComMag_2018, WiFi_assisted_60GHz_Mobicomm_2017, 2018_WCL_molish_rate_and_outage_analysis_dual_band}.
the signalling for band switching procedure by 3GPP, also referred to as a legacy approach was proposed in \cite{3gpp_36_331}. Work in \cite{Beyond5G_MAC_HWN_ComMag_2018} considered distance based thresholding to select the appropriate RF band at a software defined network (SDN) controller and proposed a medium access control (MAC) protocol for mobile heterogeneous networks. Band-switching between Sub-6 GHz and mmWave (60 GHz) in unlicensed/ISM band was experimentally studied in \cite{WiFi_assisted_60GHz_Mobicomm_2017}. Performance analysis of rate and outage probability in dual band systems with prediction based switching was proposed in \cite{2018_WCL_molish_rate_and_outage_analysis_dual_band, 2019_molish_duaband_shawdowing_correlation}, and similar approach was proposed in \cite{2022_analysis_integrated_subb6_mmWave_Globecom2017}, \cite{2020_max_system_throughput_Telkomika}.
Work in \cite{Miguel_TWC2023} analyses a hybrid transmission scheme for improved reliability
in mmWave bands with adaptive diversity combining schemes \cite{Miguel_TWC2023}. Analytical formulation of rate optimal power and bandwidth allocation in an
integrated Sub-6 GHz – mmWave architecture was proposed in \cite{2017_asilomar_rate_Shroff}.}
% Performance analysis of Integrated Sub6 and mmWave \cite{2022_analysis_integrated_subb6_mmWave_Globecom2017}, 
% %Emils work of mmasive MIMO in Sub6 and mmWave \cite{2019_emil_massiveMIMO_mmWave_Sub6},
% System throughput Dual band work in \cite{2020_max_system_throughput_Telkomika}

Recently, a few works have applied machine learning (ML) techniques in the context of band assignment. 
%For instance, handovers in dual band systems using unsupervised learning approach was carried out in \cite{2019_ML_based_handovers_in_dualband_systems_TWC}, 
% partial blind handover in dual band system using xGboost classifier was studied in \cite{DL_based_predictive_band_switching_TWC2020},
%\cite{2018_partialblind_handover_mmWave_Sub6_FarisMismar_ICCworkshop}, 
%The work in \cite{rasheed2021intelligent} mainly focuses on the design of routing protocols in SDN controllers and predictive switching between mmWave and THz bands. 
For instance, a supervised ML based approach for band switching in dual band systems was studied in 
%\cite{dual_band_learning_based_approach2018_MILCOM_Molish}
\cite{2022_bandswitch_molisch_TWC}. A deep neural network (DNN) based band switching in dual band systems was proposed in  \cite{DL_based_predictive_band_switching_TWC2020}, while deep reinforcement learning (DRL) for band switching in unmanned aerial vehicles  was analyzed in \cite{Bandswitch_DRL_UAV_hamadi_VTC2021}. 

% Although the higher frequency bands (mmWave) in multiband heterogeneous networks promises a very high data-rate, 
% %such systems are usually equipped with large number of antenna elements to overcome the high isotropic path loss. Moreover, 
% the power consumed by RF components at mmWave band is significantly higher compared to Sub-6 GHz, especially the power hungry analog to digital converters (ADCs) due to its bandwidth dependence
Most of the aforementioned works considered the criteria of maximizing the data communications rate in band assignment. Although the higher frequency bands promise a very high data rates,
the power consumption is significantly different in different bands, e.g., more than an order of magnitude higher in THz than in mmWave bands \cite{2020_6G_Summit_power_consumption_analysis_of_mmWave_SubTHz}.
%the power consumed by RF components at the mmWave band is significantly higher compared to Sub-6 GHz
% , especially the power hungry analog to digital converters (ADCs) due to its
% bandwidth dependence \cite{M_zorzi_mmWave_TWC2017}.
Since the battery power in UE is limited, it is desirable for future green networks that the band assignment policies not only consider the rate criteria but also the power consumed. To this end, a heuristic approach for dual bands (Sub-6 GHz and mmWave) was considered in our previous work \cite{Our_ccnc_paper2023} wherein if the power consumed in a recent time window exceeds a certain threshold, a band switch would be initiated to a lower-power-consuming band, without  consideration of the rate to be achieved.  

\textcolor{black}{In addition to the high power consumption associated with mmWave and emerging sub-THz bands in multiband systems, the impact of blockage must also be considered due to the sensitivity of high-frequency signals to blockage. Several prior works have investigated blockage in different contexts. For example, \cite{2020_TWC_LOS_NLOS_identification_Sub6} considered ML based techniques for LOS identification in vehicle-to-vehicle (V2V) networks at Sub-6 GHz frequencies using extensive measurement data, while \cite{2023_CL_LOS_NLOS_identification_Sub6} employed channel impulse responses and Convolutional Neural Network (CNN) models on 5G experimental data for LOS/NLOS classification. For the mmWave band, \cite{2024_adaptive_RIS_blockage_mmWave} examined blockage at 26 GHz with randomly placed obstacles and mobile UEs, mitigated adaptively using reconfigurable intelligent surfaces (RIS). Similarly, \cite{2020_Globecom_sub6_mmWave_blockage,DL_mmwavebeam_blockage_prediction_usingsub6GHZ_TCOM2020} leveraged out-of-band Sub-6 GHz information for blockage prediction in mmWave systems using deep learning on the DeepMIMO dataset. Work in \cite{2021_WCL_mmWave_Thz_linkPrediction_analytical} investigated link prediction for mmWave/THz channels but relied on simplified channel models (Rician fading) and artificial mobility patterns such as the Bernoulli lemniscate. Most of the above studies, however, address blockage within a single band (either Sub-6 GHz, mmWave, or sub-THz) and do not account for joint behavior across multiple bands. 
%To the best of our knowledge, no prior work has systematically investigated blockage in multiband networks under both outdoor and indoor scenarios. 
In contrast, this work focuses on multibands and focuses on the effects of blockage and power consumption in realistic settings.
%incorporating high-frequency challenges while explicitly modeling blockage dynamics (including Bernoulli-type mobility) across sub-6 GHz, mmWave, and sub-THz bands.
Note that in this work, we do not explicitly predict the occurrence of blockages. We generate channels that have a certain probability of having blockages, and so the different channel realizations include the effects of blockage with a certain probability.}
%Instead, we generate channels under various blockage conditions so that the channel realizations inherently capture the effects of blockage.

With the possibility of forecasting channel/rate in future time slots, enabled by advances in ML, an energy-aware BS  may choose to not utilize a high-rate, high energy consuming band (or it may choose to not transmit at all) at a given time-slot if it predicts that the rate will be significantly higher in a later time slot. Such an approach can lead to improved energy usage in systems where the highest possible rate is not always necessary. On the other hand, such an approach might lead to long delays if the BS waits for favorable channel conditions before transmitting in a high frequency, high-energy-cost band. Therefore, to make such a system practical, some form of time limitation is necessary. To the best of the authors' knowledge, such an approach which considers to minimize the power consumption and simultaneously attempting to meet the target rate in a given time constraint using rate forecast in future time slots, although promising, has not been reported in the literature.

In this context, we consider energy aware operation of downlink multiband networks. Given the battery power limitation of UE, we only consider the power consumption of the UE and not the BS.
%where time is split into frames, and frames into slots. 
We divide time into frames, and each frame into $T$ slots. In each slot of a $T$-slot frame, the BS chooses which band to transmit in to meet a target sum rate in the frame (proportional to the number of bits transmitted in the frame), with the lowest average UE power consumption in that frame. The choice of which band to utilize in this work is enabled by the use of multiple rate/channel forecasters which predict rates in future time slots based on learned past history of a moving UE, and other UEs with similar (but not the same) trajectory.  

\subsection{Contributions}
The key contributions of this work are fourfold and can be summarized as follows:
\begin{itemize}
    \item A novel framework  for 
    energy aware band assignment in multiband networks is proposed  which minimizes power consumption while attempting to  achieve a target average rate per frame.
    
    \item An iterative procedure for band assignment which utilizes rate/channel predictions for varying number of future slots is introduced. This approach can be applied to any rate/channel forecasting algorithm.
    
    %Using the acheived rate.     we take into consideration the data rate in the source band, power consumed by the RF chain at each time instant, and utilize a long short term memory (LSTM) architecture. With our proposed scheme, the median energy efficiency is 56.30\% higher as compared to the case when energy usage is not considered in the bandswitching decision, while the median rate reduces only by 13.81\%.
    % Based on the band switch request from UE, BS first predicts the target band rate
    % %Based on designed policy, BS  
    % and then decides whether to grant or deny the band switch request.
    \item \textcolor{black}{
    %To consider the impact of blockage at high frequency in multiband networks, we propose a Bernoulli blockage generation algorithm on DeepMIMO and NYUSIM outdoor settings. Moreover, in indoor settings, we developed multiband testbed in indoor setting, in addition to DeepMIMO-Indoor I3  and NYUSIM-InH scenario. 
    Considering the impact of blockage at high frequency in multiband networks, we propose a Bernoulli blockage generation model applied to DeepMIMO and NYUSIM outdoor and indoor settings. }
    %For indoor scenarios, we develop a multiband testbed in addition to employing DeepMIMO-Indoor I3 and NYUSIM-InH datasets, thereby enabling realistic evaluation under blockage conditions.
    \item  We design  rate and channel forecasters which forecast direct multistep ahead for future time slots. We propose a stacked long short term memory (LSTM) as a forecaster(s). \textcolor{black}{We tailor the publicly available ray-tracing based `DeepMIMO' \cite{deepMIMo_ahmed_dataset_gen_paper}, `NYUSIM' \cite{nyusim4} dataset for motion, and develop a multiband empirical testbed setup and apply this approach to it.}  The proposed approach outperforms several other approaches considered.
    Simulation results reveals  the LSTM forecaster also outperforms the Transformer model, which has received a lot of attention recently.
    %\textcolor{black}{The proposed scheme has an average 13.34\% higher energy efficiency than the existing band assignment policies.}  %we tailor the publicly available `DeepMIMO dataset' as per our requirement and test the robustness of the proposed policy. Moreover, we also use the `TimeSeriesSplit' approach as a validation method instead of the classical approach. %The proposed policy outperforms  the existing approaches.
\end{itemize}

\textcolor{black}{The remainder of this paper is organized as follows. Section \ref{secII} presents the network and system model, power consumption characteristics followed by the problem formulation. Section \ref{secIII} describes proposed scheme and methodology. Section \ref{secIV} focuses on dataset generation for outdoor and indoor scenarios (including blockage) using DeepMIMO, NYUSIM, and the empirical multiband testbed setup, together with the forecasting methodology. Section \ref{secV} reports and analyzes the experimental results. Finally, Section \ref{secVI} concludes the paper.}

%\textcolor{black}{The remainder of this paper is organized as follows. Section \ref{secII} presents the network and system model, along with power consumption characteristics in the Sub-6 GHz, mmWave, and THz bands for both outdoor and indoor scenarios, followed by the problem formulation. Section \ref{secIII} provides a detailed discussion of the proposed scheme and methodology. Section \ref{secIV} focuses on dataset generation for outdoor and indoor scenarios (including blockage) using DeepMIMO, NYUSIM, and the empirical multiband testbed setup, together with the forecasting methodology. Section \ref{secV} reports and analyzes the experimental results. Finally, Section \ref{secVI} concludes the paper.}

\section{System model and Problem Formulation}
\label{secII}

\subsection{Network and System Model}
In this work, we consider a  multiband network in which both the BS/access point (AP) and \textcolor{black}{ mobile UEs} can operate in the Sub-6 GHz, mmWave or sub-THz bands \textcolor{black}{(in outdoor settings), and in indoor settings under different considerations as explained in the later sections}. We assume that the BS is equipped with multiple antennas (3D array).
%of $N_{Tx} = M_x \times M_y \times M_z$).
As per the UE design in \cite{2020_6G_Summit_power_consumption_analysis_of_mmWave_SubTHz} and references therein, for UEs in outdoor scenario, we assume a single antenna at Sub-6 (3.5 GHz), an array with $N_{Rx} = 8$ antennas at mmWave (28 GHz), and an array with $N_{Rx} = 64$ antenna elements at THz band (140 GHz). \textcolor{black}{For the indoor scenario, we assume $N_{Rx} = 4$ antennas at Sub-6 (2.4 GHz), and an array of $N_{Rx} = 16$ antennas at mmWave band (60 GHz). We use array of $N_{Rx} = 64$ antennas at THz band (142 GHz) specifically in indoor NYUSIM InH and emulated sub-THz lab data, discussed later in the text.}
%We also assume that in order to maximize  the throughput and depending on the power consumption, the UE initiates the band switching procedure. 

For ease of notation, we use $i \in (\text{NoTx, Sub-6, mmWave, THz})$ to denote no-transmission case or transmission in any one of the three bands respectively, and at any time slot, the BS can only utilize one of the bands or not transmit at all. In slot $t$, the baseband equivalent of the received signal at the UE, transmitted by the BS in band $i$  can be written as \cite{DL_based_predictive_band_switching_TWC2020}:
\begin{equation}
\small
    y_i[t] = \sqrt{\text{P}_{T_{X_i}}} \hspace{1mm} \textbf{\text{h}}^*_i [t]\hspace{1mm} \textbf{\text{f}}_i [t] \hspace{1mm} x_i[t] + w_i[t],
    \label{y(t)}
\end{equation}
% \begin{equation}
% \small
%     y_i(t) = \sqrt{\text{P}_{T_{X_i}}} \hspace{1mm} \textbf{\text{h}}^*_i (t)\hspace{1mm} \textbf{\text{f}}_i (t) \hspace{1mm} x_i(t) + w_i(t),
% \end{equation}
where, $\text{P}_{T_{X_i}}$ is the transmitted power by BS in $i^\text{th}$ band, $\textbf{\text{h}}_i[t]$ $\in \mathbb{C}^{N_{Tx}}$ is the channel vector, $\textbf{\text{f}}_i [t]$ is the beam forming vector at $t$ time slot, $x_i[t]$ is the transmitted signal, and $w_i[t]$ is the additive white Gaussian noise with zero mean and variance $\sigma^2_i$ in the $i$-th band. The evolution of $\mathbf{h}_i[t]$ over  time is based on the motion of the UE and is described in detail  in Section \ref{sec:DatasetConstruction}.

\textcolor{black}{The expression in (\ref{y(t)}) is for a single antenna receiver.} Subsequently, we shall consider multiple receiver antennas and scale the resulting SNR to account for multiple receiver antennas. We assume that the BS performs maximal ratio transmission where 
$\textbf{f}_i[t]  = \frac{1}{|\textbf{h}_i[t]|}\textbf{h}_i[t]$.
%analog beamforming with the beamforming vector chosen from a predefined codebook $\mathcal{F}_i$. The BS chooses the beam forming vector $\textbf{f}^{\ast}$  that maximizes the received SNR in the $i^\text{th}$ band at UE. It can be written as:
% \begin{equation}
%     \text{\textbf{f}}^{\ast}_i:= \hspace{1mm} _{\text{\textbf{f}}_i \in \mathcal{F}_i}^{\text{arg max}} |\text{\textbf{h}}^*_i \hspace{1mm} \text{\textbf{f}}_i|^2.
% \end{equation}
\textcolor{black}{While the datasets do not provide MIMO channel matrices, we simulate the effect of multiple receive antennas by assuming that the SNR is scaled linearly with the number of receiver antennas.} Accordingly, the received SNR on the $i^\text{th}$ frequency band with $N_{Rx}$ receive antennas is
\begin{equation}
\small
    \gamma_i[t] = \frac{\text{P}_{T_{X_i}}}{\sigma_i^2} N_{Rx} \hspace{1mm} |\text{\textbf{h}}_i[t]|^2.
    \label{snr_eqn}
\end{equation}
Note that we have made the assumption that the SNR with multiple receive antennas is equal to the SNR with one receive antenna multiplied by the number of antennas. This assumption enables us to easily use datasets generated for UEs with isotropic antennas. 
The achievable rate in slot $t$ is:
\begin{equation}
\small
    R_i[t] = B_i \hspace{1mm}\text{log}_2\{1 +  \gamma_i[t]\},
    \label{rate_eqn}
\end{equation}
where $B_i$ is the bandwidth of the $i^\text{th}$ frequency band, with $B_{\text{NoTx}} = 0$. \textcolor{black}{For outdoor settings, we assume $B_{\text{sub-6}}=10 \text{ MHz}$, $B_{\text{mmWave}}=100\text{ MHz}$, and $B_{\text{THz}}=1\text{ GHz}$, while for the indoor settings $B_{\text{sub-6}}=80\text{ MHz}$, $B_{\text{mmWave}}=1\text{ GHz}$, and $B_{\text{sub-THz}}=3\text{ GHz}$.} %\textcolor{black}{Briefly bout coherence times.} 
We assume that the BS is equipped with a forecaster and learns from past history of other UEs to make energy aware band assignments\footnote{The proposed scheme is generic, but is consistent with, and can also easily be extended to next generation of RAN architectures wherein RAN intelligent controllers (RICs) use analytics and drive the network actions.}.
\begin{table}[t!]
\centering
\caption{Power consumption of RF components (in mW) for outdoor and indoor scenarios.}
\resizebox{\linewidth}{!}{%
\begin{tabular}{|c||c|c|c||c|c|}
\hline
\multirow{2}{*}{\vspace{-0.35cm}\textbf{Component}} & \multicolumn{3}{c||}{\textbf{Outdoor Scenario}} & \multicolumn{2}{c|}{\textbf{Indoor Scenario}} \\ \cline{2-6} 
 & \begin{tabular}[c]{@{}c@{}}\textbf{Sub-6}\\ \textbf{(3.5GHz)}\end{tabular} & 
   \begin{tabular}[c]{@{}c@{}}\textbf{mmWave}\\ \textbf{(28GHz)}\end{tabular} & 
   \begin{tabular}[c]{@{}c@{}}\textbf{THz}\\ \textbf{(140GHz)}\end{tabular} & 
   \begin{tabular}[c]{@{}c@{}}\textbf{Sub-6}\\ \textbf{(2.4GHz)}\end{tabular} & 
   \begin{tabular}[c]{@{}c@{}}\textbf{mmWave}\\ \textbf{(60GHz)}\end{tabular} \\ \hline
Bandpass filter (\texttt{BPF})              & 5     & 5     & 5      & 5     & 5     \\ \hline
Low Noise Amplifier (\texttt{LNA})          & 10    & 11.13 & 50.89  & 10    & \textcolor{black}{39} \\ \hline
Local Oscillator (\texttt{LO})              & 5     & 5     & 5      & 5     & 5     \\ \hline
Phase Shifter (\texttt{PS})                 & -     & 1.5   & 1.5    & -     & 2     \\ \hline
Combiner                                    & -     & 19.5  & 19.5   & -     & 19.5  \\ \hline
Mixer                                       & 15    & 16.8  & 49     & 15    & 16.8  \\ \hline
Low Pass Filter (\texttt{LPF})              & 10    & 14    & 11.36  & 10    & 14    \\ \hline
Baseband Amplifier (\texttt{BBA})           & 5     & 5     & 5      & 5     & 5     \\ \hline
Analog to Digital Converter (\texttt{ADC})  & 7.8   & 8.2   & 32.7   & \textcolor{black}{1.33} & \textcolor{black}{35.9} \\ \hline
\end{tabular}}
\label{rf_power_table_merged}
\end{table}
\subsection{Power Consumption in Sub-6, mmWave, and THz band}
Since the UEs can operate in Sub-6, mmWave, or the THz bands, the power consumed by the RF chains will be significantly different. The estimates of power consumed by the major components in the RF chains  is summarized in Table \ref{rf_power_table_merged}, and can be expressed as:
%The power consumed by the RF chain 
\begin{equation}
\small
\begin{split}
    \text{P}_{i} &= N_{i-Rx}(\text{P}_{i-\texttt{BPF}}+\text{P}_{i-\texttt{LNA}} + \text{P}_{i-\texttt{PS}}) +\text{P}_{i-\texttt{Combiner}}+\text{P}_{i-\texttt{LO}}+ \\\hspace{0.7cm}& 2(\text{P}_{i-\texttt{Mixer}}+\text{P}_{i-\texttt{LPF}} +\text{P}_{i-\texttt{BBA}}+\text{P}_{i-\texttt{ADC}}).
\end{split}
\label{Power_eqn}
\end{equation}
The factor-2 in \eqref{Power_eqn} is due to the inphase and quadrature phase components. 
%We would like to highlight that although the phase shifter and combiner are present in mmWave receiver, they are not considered since UE is assumed to be equipped with single antenna. 
On substituting the values of each component in (\ref{Power_eqn}) based on \cite{2020_6G_Summit_power_consumption_analysis_of_mmWave_SubTHz, Energy_constraint_sub6_goldsmith_TWC2005,system_level_Energy_sub6_TVLSI2007, M_zorzi_mmWave_TWC2017}  and references therein, the approximate power consumed \textcolor{black}{(in outdoor settings)} by the RF chain of the UE  in Sub-6 GHz, mmWave and THz bands are  $\text{P}_{\text{Sub-6}}=85.60$ mW, $\text{P}_{\text{mmWave}}=254.90$ mW and $\text{P}_{\text{THz}}=3893.58$ mW,  respectively. 
%\textcolor{black}{(Include a brief statement about the transceiver,later after 1st draft, as stated in proposal.)} 
Note that the power consumption is more than an order of magnitude higher in the THz than in the mmWave and Sub-6 GHz bands. \textcolor{black}{Similarly, in indoor settings, $\text{P}_{\text{Sub-6}}=82.66$ mW, $\text{P}_{\text{mmWave}}=535.90$ mW, while the power consumption for the sub-THz is assumed to be equivalent to that observed in the outdoor THz.} \textcolor{black}{We further note that practical measurements of power consumption of high frequency systems such as mmWave systems was found to be a lot higher than the calculations based on power consumption of individual components due to additional power hungry features of the high-frequency bands such as the increased power used for beam steering, as reported in \cite{saha2017detailed} which measured power consumption of $\sim 2$ W for 60 GHz mmWave systems.}

Our aim is to design an energy aware band assignment approach which minimizes the power consumption subject to a target sum/average rate per frame, which is mathematically formulated in the next subsection.

\subsection{Problem Formulation}
%\textcolor{red}{Modify the problem statement for generic N bands.}
To minimize the energy usage, one option is to never transmit and thus power consumption would be minimal, which is unrealistic. Thus, we introduce a target rate that the BS  will try to achieve at the UE.  Further, in trying to achieve a sum target rate in a manner that is energy efficient, the BS may wait for a long time for favorable channel conditions before transmitting at all, or before  transmitting in a high frequency, high energy-cost band. Such an approach could lead to extremely long delays. Thus, we introduce a delay constraint $T$. We assume a time slotted communication, with a frame comprising of  $T$ time slots.
\textcolor{black}{The aim of the BS is to minimize the power consumed in every frame while aiming to achieve a target sum-rate of at least $M$ bits/sec in that frame at the UE, as opposed to minimizing only energy per bit.}
%The aim of the BS is to achieve a target sum-rate \textcolor{black}{of at least $M$ bits/sec} in every frame while utilizing the lowest average power during that frame at the UE, \textcolor{black}{as opposed to minimizing only energy per bit}.
Further, we assume that the BS  can choose which band to transmit in for each slot, or choose not to transmit at all in a given slot. 
% This can be expressed with the aid of eq. (\ref{rate_eqn}) as: 
%  %(highlight term I and II in the equation):
% \begin{equation}
%     M \leq  \sum_{t=1}^{T} B_{i_{t}} \hspace{1mm}\text{log}_2\left(1 +  \frac{\textcolor{red}{\text{P}_{i}} \hspace{1mm}|\text{\textbf{h}}_{i_t}|^2}{\sigma_{i_t}^2}    \right),
%     \label{M_based_eqn}
% \end{equation}
% where $|\text{\textbf{h}}_i|^2 = |\text{\textbf{h}}^*_i[t] \hspace{1mm} \text{\textbf{f}}_i[t]|^2$, \textcolor{red}{$\text{P}_{i}$ is the UEs transmit power} in the  $i^{th}$ band. 
With the aid of eq. (\ref{rate_eqn}), for a given frame with $T$ slots and $i_t$ representing the band in the $t$-th slot during the frame,  we wish to find $i_t$'s to
% \begin{align}
% %\small
% &\text{min.} \hspace{2mm} \text{P}_{\text{consumed}} = \sum_{t=1}^{T} {\text{P}_{i_t}} ,    \notag \\
%  {\rm s}.{\rm t}. \hspace{1mm}  M &\leq  \sum_{t=1}^{T} \left(1-\nu \mathbbm{1}_{\{i_t\neq i_{t-1}\}}\right)B_{i_{t}} \hspace{1mm}\text{log}_2\left(1 +  \underbrace{\frac{\text{P}_{\text{Tx}_{i_t}} \hspace{1mm}|\text{\textbf{h}}_{i_t}[t]|^2}{\sigma_{i_t}^2}}_{\gamma}    \right), \notag \\
%  \textcolor{black}{\text{P}_{i_t}}  &\geq 0 
%     \label{M_based_eqn}
% \end{align}
\begin{subequations}\label{eq:power_opt}
\begin{align}
\small
\text{min.} \quad 
& \text{P}_{\text{consumed}} = \sum_{t=1}^{T} \text{P}_{i_t}, \label{eq:power_opt_a} \\
\text{s.t.} \quad 
M &\leq \sum_{t=1}^{T} \left(1-\nu \mathbbm{1}_{\{i_t\neq i_{t-1}\}}\right)
B_{i_t} \log_2\!\left(1 + 
\underbrace{\frac{\text{P}_{\text{Tx}_{i_t}} |\mathbf{h}_{i_t}[t]|^2}
{\sigma_{i_t}^2}}_{\gamma}
\right), \label{eq:power_opt_b} \\
\textcolor{black}{\text{P}_{i_t}} 
&\geq 0, \label{eq:power_opt_c}
\end{align}
\end{subequations}
%\begin{equation}
%\small
%\begin{split}
%    M &\leq  \sum_{t=1}^{T} \left(1-\nu 1_{\{i_t\neq %i_{t-1}\}}\right)B_{i_{t}} %\hspace{1mm}\text{log}_2\left(1 +  %\frac{\text{P}_{\text{Tx}_{i_t}} %\hspace{1mm}|\text{\textbf{h}}_{i_t}|^2}{\sigma_{i_t}^2}    \right), \\ &
%    {\rm s}.{\rm t}. \qquad    \text{min.} \hspace{2mm} %\text{P}_{\text{consumed}} = \sum_{t=1}^{T} {\text{P}_{i_t}}
%    \label{M_based_eqn}
%    \end{split}
%\end{equation}
where $\text{P}_{T_{X_i}}$ is the BS transmit power in the  $i^{th}$ band, $\nu \in [0,1)$  is a switching cost which models the overhead time required to switch bands, and $\mathbbm{1}$ is an indicator function.
%\textcolor{black}{that equals 1 if $i_t \neq i_{t-1}$, and 0 otherwise}. 
Here, we show the equations just for a single frame to simplify notation. Our objective is to minimize the average consumed power 
in each frame subject to the target average rate\footnote{We interchangeably use the terms target rate and rate threshold.} in that frame.
In other words, we attempt to minimize $\sum_{t = 1}^T P_{i_t}$
subject to \eqref{eq:power_opt_b} and \eqref{eq:power_opt_c}.  If \eqref{eq:power_opt_a} is not achievable, then we aim to maximize the \textcolor{black}{right side} of \eqref{eq:power_opt_b}.

Note that if channel coefficients/rates of future time slots were known,  the optimal channel assignment can be done by exhaustive search. Since future channels cannot be known at a prior time slot, the BS predicts the rate/channel conditions in future time slots to make its decision on which band to use in the current time slot. \textcolor{black}{We would like to highlight that since the prediction will be done at the BS/AP as it will be able to utilize the learned history of previous UEs, the computational power cost will not be a significant factor.}
%$B_i$ is the bandwidth of the $i^\text{th}$ frequency band and $\sigma^2$ is the noise variance. 
%In each slot, the BS-UE pair chooses which channel to transmit in to meet a target average rate in the frame, with the lowest average power consumption in that frame. 
In this work, the choice of which band to transmit in is enabled by the use of multiple rate forecasters which predict rates in future time slots based on learned past history, which is discussed in the next section.
We would like to highlight that the rate in eq. \eqref{eq:power_opt_b} can be forecasted by: \textbf{1)} forecasting the $\text{log}_2\left(1 +  \gamma \right)$ term, which we refer to as rate forecaster\footnote{With the abuse of notation, we refer spectral efficiency term as rate.}, or  \textbf{2)} By forecasting the $|\text{\textbf{h}}_i[t]|^2$, which we refer to as channel forecaster. 

% \textcolor{red}{Write 1-2 line about why analytical approach would fail here.}
% \textcolor{black}{Also need to mention somewhere that the proposed scheme is proposed does not send a request to the base station, instead base station intelligently forecasts and take the decision by itself. Also include about the measurement gap in one or two statement.}
% = 30.3+50+20+2.5+3+0.14 = 105.5 \text{mW} Similarly, the power consumed by UE in mmWave band can be given as:
% \begin{equation}
% \begin{split}
%     \text{P}_{\texttt{mmWave}} &= N_r(\text{P}_{\texttt{LNA}}+\text{P}_{\texttt{PS}})+\text{P}_{\texttt{RF}}+\text{P}_{\texttt{Combiner}}+2\text{P}_{\texttt{ADC}}\\
%     &= 1(39+2)+40.8+19.5+2(62.093)
%     = 225.486 \text{mW}
% \end{split}
% \end{equation}
% where, $\text{P}_{\texttt{RF}} = \text{P}_{\texttt{Mixer}}+\text{P}_{\texttt{LO}}+\text{P}_{\texttt{LPF}}+\text{P}_{\texttt{Baseband amp.}}$

\section{Proposed Scheme \& Methodology}
\label{secIII}

%\subsection{Proposed scheme}
\begin{figure}[t!]
\centering
\includegraphics[width=3.7in, height = 4.5cm]{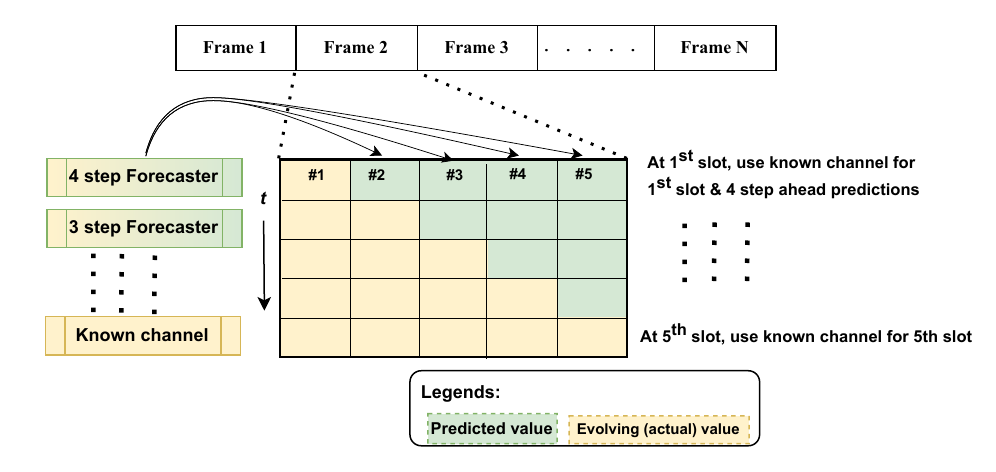}
\caption{Framework of the proposed scheme.}
\label{proposed_scheme_figure}
\end{figure}
\begin{algorithm}[t!]
\caption{Proposed algorithm}
\label{proposed_alogrithm}
\begin{algorithmic}[1]

\Procedure{Proposed Approach}{}
\State Import scenario; select UEs, BS (or AP) locations
    %\State Select Locations of UE and BS 
    \State Run ray traces for the settings specified
    %\State Divide UE in clusters
    \State Specify $M$, Power consumed in each band
\State Built and model $T$ separate forecasters.
\State  Each forecaster predicts $T$ continuous predictions
\State Load the forecaster values
\For {$m=1$ : \texttt{all frames}}
    \For {$t=1$ : $T$}
        \State Obtain $R_i[t]$ for $i=0:K$ \{\text 0 = No Tx, 1 = Sub-6, 2 = mmWave, 3 = THz, \ldots\}
    \State Get $\hat R_i[t+1,t]... \hat R_i[T,t]$, {\small $i= \{0, 1, 2, 3\}$}
%    \State Multiply forecasted values by $0.85, 0.8,\cdots$
     \State $\hat R(i_t, i_{t+1},\cdots i_T) \leftarrow R_{i_{t}} [t]+\sum_{k = t+1}^T\hat R_{i_{k}}[k],$\\
     \hspace{1.7cm}$ \forall (i_t, \cdots i_T)\in \{\text{\small  NoTx,Sub-6,mmWave,THz}\}^{T-t+1}$
\State Call Optimal BandAssignment with \\
\hspace{1.8cm}$\hat R(i_t, i_{t+1},\cdots i_T), M, T-t+1$
%\State $(\hat i_t, \cdots, \hat i_T) \leftarrow \argmin_{(i_t, \cdots i_T)} 
%P_{i_t}+\cdots+P_{i_T}$, \\
%\hspace{1.7cm} s.t. $R(i_t, i_{t+1},\cdots i_T)\geq M$
\State $b_t \leftarrow $ slot $t$ assignment by Opt. BandAssignment
    \State $M\leftarrow M-R_{b_t}[t]$
    \EndFor
    \State Compute  power consumed in frame.
    \State Compute  sum rate in frame.
    \EndFor
\State Compute average power consumed per frame.
\EndProcedure
\end{algorithmic}
\end{algorithm}
\begin{figure*}[t!]
    \centering
    \captionsetup{format =hang}%
   % \vskip\baselineskip
    \begin{subfigure}[t]{0.49\textwidth}  
        \centering 
        \includegraphics[scale=0.7]{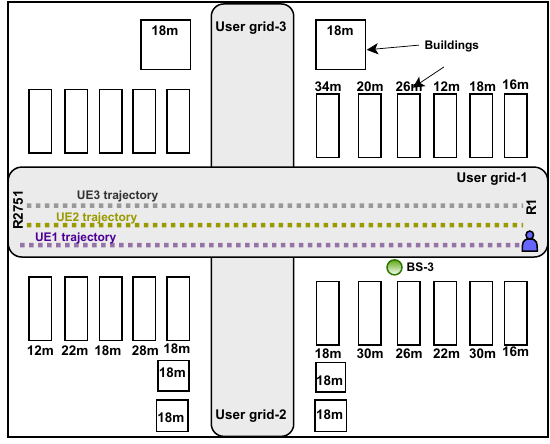}
        \caption[]%
        {\small{Top view of DeepMIMO `O1\_outdoor scenario' \cite{deepMIMo_ahmed_dataset_gen_paper}}}    
        \label{O1}
    \end{subfigure}
%    \hspace{0.1cm}
%    \quad
    %\hfill
    \begin{subfigure}[t]{0.49\textwidth}  
        \centering 
        \includegraphics[scale=0.55]{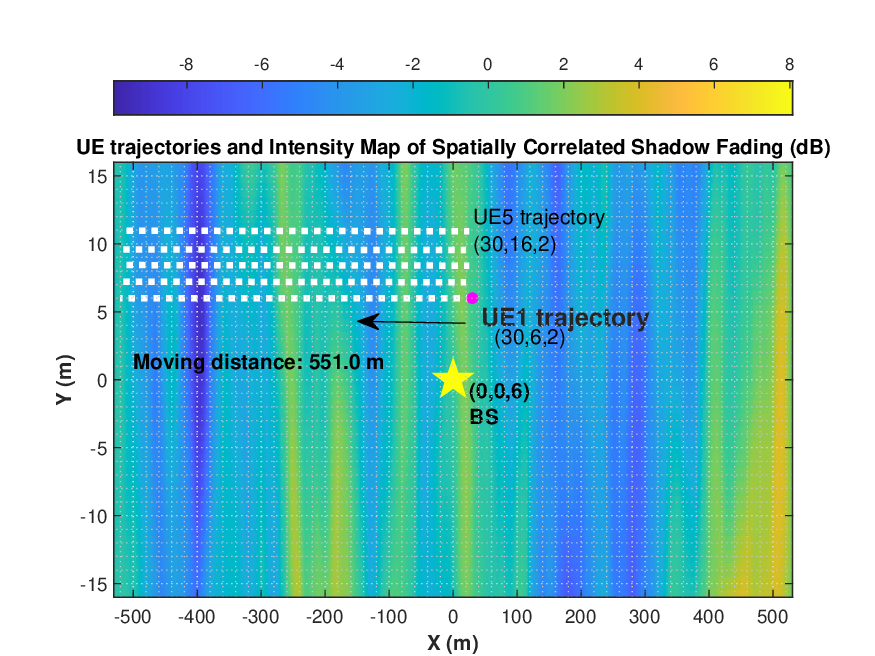}
        \caption[]%
        {{\small NYUSIM Outdoor UMi scenario \cite{nyusim4}}}    
        \label{nyusim_fig}
    \end{subfigure}
%    \quad 
    \caption{UE trajectories in outdoor scenarios. Figures not to scale.}
    \label{UE_trajec_images}
\end{figure*}
%Different from the existing approaches, we consider the rate and power criteria in designing an energy efficient band assignment scheme. 
\textcolor{black}{The propagation characteristics of wireless signals can vary significantly across frequency bands. For example, signal attenuation, interference levels, and power consumption at 2.4 GHz can be quite different from those at 5 GHz within the Sub-6 GHz band. Similarly, at higher frequencies such as 24 GHz compared to 60 GHz and 140 GHz, the propagation conditions and wireless channel behavior differ considerably. To address this, we propose a scheme that generalizes across $`K$' bands, making it applicable to a wide range of operating frequencies.}
The framework of the proposed scheme is as shown in Fig. \ref{proposed_scheme_figure} and complemented with algorithm-1 \& 2. Assume that each frame is comprised of $T=5$ slots for illustration (any number $T> 1$ could work). \textcolor{black}{ For the sake of clarity and tractability, in this work we restrict our analysis to three bands, as discussed earlier. At each time, the BS (or AP in case of indoor) can either not transmit i.e., $(K = 0)$
%\footnote{This is similar to the discontinuous reception for energy efficient communication \cite{discontinuous_reception_survey}.}, 
or utilize one of the  Sub-6, mmWave, or  THz bands $(K = 1, 2, 3)$ respectively.} Thus, there are total four options i.e., no transmit/Sub-6/mmWave/THz, of which one is to be selected in each time slot.  For better understanding, a grid structure is shown in Fig. \ref{proposed_scheme_figure} wherein the rows are the time indices, and columns are the slots. We propose to use `$T$'  forecasters, each  predicting `$T$' direct multistep ahead i.e., 4 step forecaster predicts 4 direct step ahead continuous values. Thus, in the first slot, the 4 step forecaster would be used to predict the rate/channel values for all the future time slots in the frame, where we assume that the rate/channel in the current slot is known.   The forecasted values are highlighted by the green boxes in the first row of the grid. Since rate/channel predictions for future slots become less reliable (particularly for THz), we discount the channel predictions further into the future to account for this fact. We do so by multiplying rates/channels for the THz forecast by factors of 0.85, 0.8, 0.75, and 0.7 for the predictions 1 - 4 steps into the future respectively (algorithm-1, upto line 13). Next, the optimal band assignment is done for all the five slots, using an exhaustive search.  

\begin{algorithm}[t!]
\caption{Optimal Band Assignment algorithm}
\label{optimal_bandassignment_alogrithm}
\begin{algorithmic}[1]
\Procedure{\textcolor{black}{\small Optimal BandAssignment}}{$ \hat R(i_t,\cdots i_T),M,T$}
%    \For{$m=1$ : \texttt{all frames}} %\Comment{Loop over all frames}
        % \State Read the channels/rates for this frame;
        \For{$k =1:4^T$ } % \Comment{$T=5$ in our case}
         \State Loop over all the possible band combinations
         \State Compute corresponding sum rates ($R$) 
         \State Save the  band combinations where $R>$  $M$
         \EndFor
         \If{Indices where $R>M \neq \emptyset$ }
         \State Return band assignments where energy is min.
         \Else
         \State Return band assignments where rate is max.
         \EndIf
          %\State Loop over all the frames.
       % \EndFor
    %\State This yields best assignment that satisfies objective (\ref{M_based_eqn})
\EndProcedure 
\State The above procedure yields the  best band assignments for each frame  that satisfies our objective in \eqref{eq:power_opt_a}.
\end{algorithmic}
\end{algorithm}
This is shown in algorithm-2. For each frame, it iterates over all the possible combinations i.e., $4^T$ options of band assignments (plus the no-transmit option), to obtain the set of 5 band assignments which consumes the minimum energy, and is predicted to meet the target rate. The resulting band assignment for the first slot is selected as the band to use in this slot (in algorithm-1).  The rate achieved in this slot is computed and subtracted from the sum-rate threshold (algorithm-1, line 17) to compute a new threshold for the next time slot. 
%The band assigned to slot 1 will be used for this slot. 
%Based on the optimal band assignments, Rate R would be computed. This would be the common step for the 1st time step in each frame. 
Next, at the $2^{nd}$ time slot,  we use the exact channel in the second slot and 3 step ahead predictions and so on. 
%4 step forecaster predicts four step ahead continuous values, as depicted in the 2nd row. 
At each time slot, the optimal band assignment procedure (i.e. algorithm-2) would be called with successively smaller number of slots to reflect the declining number of slots remaining in the current frame. The rate threshold \textcolor{black}{would be} updated \textcolor{black}{at} each slot by subtracting the rate already achieved in the previous time slot(s).  This process will be repeated for all the slots. The proposed algorithm is described in algorithm-1, where for a given frame, $\hat R_i[\ell,t]$ is the predicted rate for the $\ell$-th slot of the $i$-th band, as predicted in the $t$-th slot.  The rate forecasters predict  $\hat R_i[\ell,t]$ directly, while the channel forecasters use \eqref{snr_eqn} and \eqref{rate_eqn} with the predictions of $|\mathbf{h}_i[t]|^2$. 
%, \textcolor{black}{where it is desirable to utilize a lower power band if the desired target rate is already achieved.}
%\textbf{a. Current channel prediction:}
%\textcolor{black}{One variation of the proposed scheme is that we simply use the rate in the current slot as the forecast for the remaining slots in the frame. We refer to this as \textbf{current channel prediction} approach. Since the choice of which band to transmit in at any given slot is influenced by the achievable rates in future slots in the current frame, accordingly in this approach, the rates/channels in all bands are estimated at the start of each slot.  In deciding which band to transmit during the current slot, the BS assumes that the remaining slots in the frame will have identical rates/channels to the rates/channels estimated in this slot. Through exhaustive search, the BS computes the band assignment for the remaining slots in the frame in order to minimize power consumption and meet the target rate. From the obtained band assignments, it selects the band assigned to its current slot and uses that band. It then subtracts the achieved rate in this slot from the target sum rate for the frame in preparation for the next time slot.}
The proposed design serves as an energy aware band assignment scheme which is motivated by systems with a finite battery life, which takes advantage of the fact that rate/channel forecasts for slots longer in the future will be less accurate than forecasts for slots closer in the future.
\section{Dataset generation and  Forecasting Methodology}
\label{secIV}
\subsection{Dataset Generation and Processing: Outdoor Scenario} \label{sec:DatasetConstruction}
\subsubsection{DeepMIMO O1 dataset (LOS)}
% \begin{figure}[t!]
% \centering
% \includegraphics[scale=0.7]{results/globecom_fig.pdf}
% \caption{\small{Top view of `O1\_outdoor scenario' \cite{deepMIMo_ahmed_dataset_gen_paper}. Not to scale. }}
% \label{O1}
% \end{figure}

\begin{table}[t!]
\centering
\caption{DeepMIMO dataset and wireless parameters}
\renewcommand{\arraystretch}{1.15}
\begin{tabular}{|c|c|}
\hline
\textbf{Parameter} &  \textbf{Value} \\ \hline
%Active BS & 3 \\ \hline
% Active UE & 1 (include more details)\\ \hline
\# of BS antennas $(M_x \times M_y \times M_z)$ & (1$\times$ 8 $\times$ 4) \\ \hline
\# of UE antennas \{Sub-6; mmWave; THz\} & \{1; 8; 64\} \\ \hline
% \# of channel paths (strongest gets selected) & 3 \\ \hline
%Subcarrier Bandwidth \{Sub-6 GHz and mmWave\} & \{180, 1800\}KHz \\ \hline
$\text{P}_{T_{X_i}}$ ; Cyclic prefix ratio ; \# of channel paths & 1W ; 64 ; 3\\ \hline
$f_c$ \{Sub-6; mmWave; THz\} & \{3.5; 28; 140\} GHz \\ \hline
Bandwidth \{Sub-6; mmWave; THz\} & \{10; 100; 1000\} MHz \\ \hline
\# of OFDM subcarriers; UE speed ; $(v_s)$ & 64 ; 10 m/s (36 kmph) \\ \hline
%Cyclic prefix ratio & 0.125 \\ \hline
%UE speed $(v_s)$ & 10 m/s (36 kmph) \\ \hline
% $T_{beam}$ & 10 m/s (36 kmph) \\ \hline
% $\alpha$ & 10 m/s (36 kmph) \\ \hline
% $\theta$ & 10 m/s (36 kmph) \\ \hline
\end{tabular}
\label{deepmimo_parameters}
\end{table}
The proposed framework was evaluated on the  `DeepMIMO' dataset \cite{deepMIMo_ahmed_dataset_gen_paper}. In particular, we consider the ray-tracing outdoor scenario-1 `O1\_3p5' for Sub-6 GHz band, `O1\_28' for mmWave band, and `O1\_140' for THz band. 
The design is consistent with our considered system model.
%Since the mmWave is more prone to blockage, we also consider the  `O1\_blockage scenario' to replicate the real world scenario in simulation framework. This is shown in Fig. \ref{O1_blockage}. We combine the channels in blockage and non-blockage 28GHz scenarios using a Bernoulli random variable with a blockage probability as $0.2$. 
For all the ray traces, we consider `User grid-1' of the dataset and one active BS (i.e., BS-3) as shown in Fig. \ref{O1}. The coordinates of BS-3 is (235.50, 489.50, 6) m.  In the original dataset (also shown in Fig. \ref{O1}), the user grid-1 includes 2751 rows, R1 to R2751, with each row separated by 0.2 m. In the original dataset, there are 181 users per row. However, we sample the users and consider three active users per each row, to mimic three UEs
%\footnote{This can be easily extended to even more UEs.} 
moving along the way, each separated by 3m.  
The initial coordinates of UE1 is (242.42, 297.17, 2) m and the final coordinates are (242.42, 847.17,	2) m. Thus, there are total 2751 location datapoints per UE. This setting is equivalent to a UEs driving alongside the road with sampled rows being a time instances as indicated by the UE trajectory in Fig. \ref{O1}. The key idea for adapting this approach is that the forecaster at the BS can learn about the rates and/or channel strengths of the UEs which were in the close proximity (UE2 and UE3 in this case) in the past, and leverage this information for forecasting, the details of which are provided in section IV-D.
Table \ref{deepmimo_parameters} summarizes DeepMIMO O1 dataset parameters.
%The list of DeepMIMO dataset parameters is summarized in Table \ref{deepmimo_parameters}.

%\textcolor{black}{Due to i) the unavailability of blockage scenario at THz frequencies in DeepMIMO, and ii) to validate our findings on different dataset not only in the outdoor but also in the indoor scenario, we explored publicly available NYUSIM dataset to test our proposed algorithms, as described in the next sub-section.}

\subsubsection{NYUSIM UMi dataset (LOS and NLOS)}

\textcolor{black}{Due to the lack of blockage scenarios at THz frequencies in DeepMIMO, and to further validate the proposed approach on a different dataset, we also evaluate the algorithm using the MATLAB-based NYU Channel Model Simulator (NYUSIM 4.0) \cite{nyusim4} in spatial consistency mode. NYUSIM is based on extensive real-world propagation measurements over 0.5–150 GHz. For the outdoor case, we consider the Urban Microcell (UMi) scenario, tailored to resemble the DeepMIMO setup and remain consistent with our system model. As shown in Fig. \ref{nyusim_fig}, five users are placed along a row, each separated by 2.5 m and moving over a distance of 551 m. This yields 1600 data points per user. The initial coordinates of UE1 are (30, 6, 2) m, and the BS coordinates are (0, 0, 6) m.}

\textcolor{black}{In NYUSIM, both line-of-sight (LOS) and entirely non-LOS (NLOS) scenarios can be generated. In practice, however, users may alternate between LOS and NLOS, both indoors and outdoors. To capture this behavior, we construct a blockage-segment dataset as outlined in algorithm-\ref{blockage_alogrithm}. First, LOS-only and NLOS-only datasets are generated for all five users. Based on the blockage probability 
$(p)$ and a Bernoulli-based segment size, we compute the Bernoulli blockage \textbf{b} spanning the full trajectory. We then combine \textbf{b} with the channel realizations \textbf{h} as in line 8 of algorithm-3. At each time slot, given probability $(p)$, blockage segments of length 15 are generated. For evaluation, we set $p = 0$ for LOS, $p = 0.25$ for low blockage, $p = 0.5$ for moderate blockage, $p = 0.75$ for high blockage, and $p = 1$ for the fully NLOS extreme case.}

%Since the mmWave is more prone to blockage, we also consider the  `O1\_blockage scenario' to replicate the real world scenario in simulation framework. This is shown in Fig. \ref{O1_blockage}. We combine the channels in blockage and non-blockage 28GHz scenarios using a Bernoulli random variable with a blockage probability as $0.2$. 
%This setting is equivalent to a UEs driving alongside the road with sampled rows being a time instances as indicated by the UE trajectory in Fig. \ref{O1}. The key idea for adapting this approach is that the forecaster at the BS can learn about the rates and/or channel strengths of the UEs which were in the close proximity (UE2 and UE3 in this case) in the past, and leverage this information for forecasting, the details of which are provided in the next subsection. The list of DeepMIMO dataset parameters is summarized in Table \ref{deepmimo_parameters}.
\begin{algorithm}[t!]
\caption{\textcolor{black}{Algorithm for blockage segments generation}}
\label{blockage_alogrithm}
\begin{algorithmic}[1]
%\Procedure{\textcolor{black}{\small Generate\_blockage}}{$ modify param$}
\State Generate LOS, NLOS scenarios, and corresponding $\textbf{h}$
\State Define seed value \Comment For repeatability
\State Define blockage probability $(p)$, and Bernoulli segment length $(s)$
\State Bernoulli\_segment = np.random.binomial(1, $p$, $s$)
\State Blockage repeat = (\texttt{all coordinates} / $s$) times
\State \textbf{b} =  Bernoulli\_segment $\times$ Blockage repeat \Comment{Bernoulli blockage}
%    \For{$m=1$ : \texttt{all frames}} %\Comment{Loop over all frames}
        % \State Read the channels/rates for this frame;
        \For{$i =1 :$ \texttt{all coordinates} } % \Comment{$T=5$ in our case}
         %\State $b_i$ = Bernoulli$(p)$
         \State $\textbf{h}_i$ = $\textbf{b}$ $\cdot$ ${\textbf{h}_i}_\text{NLOS}$ + (1 - $\textbf{b}$)  $\cdot$ ${\textbf{h}_i}_\text{LOS}$
         \EndFor
          %\State Loop over all the frames.
       % \EndFor
    %\State This yields best assignment that satisfies objective (\ref{M_based_eqn})
%\EndProcedure 
\State The above algorithm yields the quantifiable blockage segments generation. 
\end{algorithmic}
\end{algorithm}

\begin{figure*}[t!]
\centering
\includegraphics[scale=0.41]{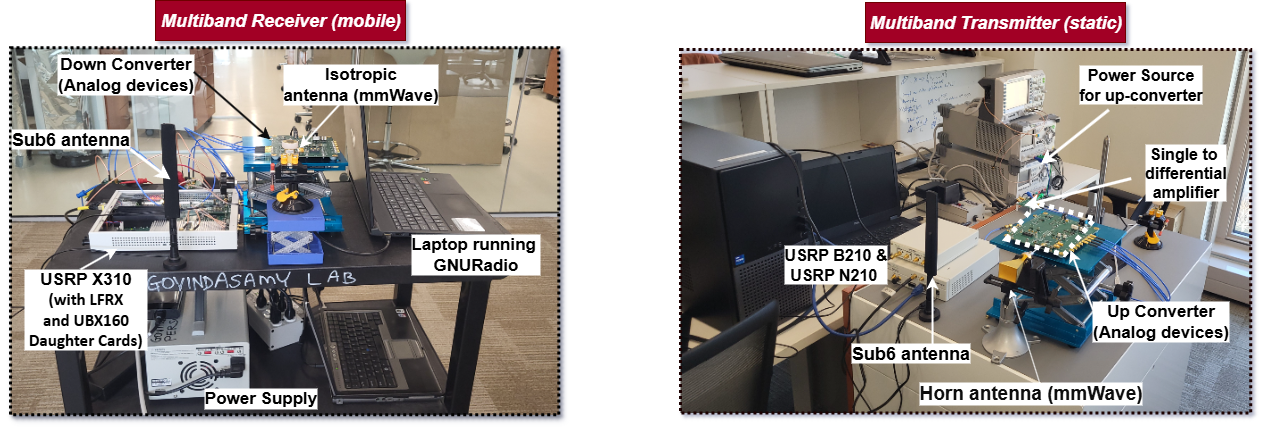}
\caption{\small{Multiband Empirical Testbed setup (Indoor Scenario)}}
\label{multiband_setup}
\end{figure*}

\begin{table*}[t!]
\centering
\caption{\textcolor{black}{Summary of sequences used in forecasting for different datasets}}
\renewcommand{\arraystretch}{1.2}
%\resizebox{\textwidth}{!}{%
\begin{tabular}{|c|cc|ccc|}
\hline
\multirow{2}{*}{\vspace{-5mm}\textbf{No. of sequences $(\downarrow)$ \textbackslash{}\textbackslash{} Scenario type$(\to)$}}
& \multicolumn{2}{c|}{\textbf{Outdoor scenario}} & \multicolumn{3}{c|}{\textbf{Indoor scenario}} \\ \cline{2-6} 
 & \multicolumn{1}{c|}{\textbf{\begin{tabular}[c]{@{}c@{}}DeepMIMO\\ O1 dataset\end{tabular}}} & \textbf{\begin{tabular}[c]{@{}c@{}}NYUSIM\\ UMi dataset\end{tabular}} & \multicolumn{1}{c|}{\textbf{\begin{tabular}[c]{@{}c@{}}DeepMIMO\\ I3 dataset\end{tabular}}} & \multicolumn{1}{c|}{\textbf{\begin{tabular}[c]{@{}c@{}}NYUSIM\\ InH dataset\end{tabular}}} & \textbf{\begin{tabular}[c]{@{}c@{}}Multiband testbed\\ dataset\end{tabular}} \\ \hline
No. of training sequences & \multicolumn{1}{c|}{6000} & 6000 & \multicolumn{1}{c|}{8785} & \multicolumn{1}{c|}{8785} & 8000 \\ \hline
No. of validation sequences & \multicolumn{1}{c|}{500} & 500 & \multicolumn{1}{c|}{1115} & \multicolumn{1}{c|}{1000} & 950 \\ \hline
No. of test sequences & \multicolumn{1}{c|}{1150} & 1150 & \multicolumn{1}{c|}{1095} & \multicolumn{1}{c|}{1095} & 1095 \\ \hline
\end{tabular}%
\label{summary_of_sequences}
\end{table*}

\subsection{Dataset Generation and Processing: Indoor Scenario}
\subsubsection{DeepMIMO I3 dataset (LOS and NLOS)} %\textcolor{black}{We validate proposed algorithm in indoor settings. \textcolor{red}{Due to the availability of datasets, we restrict to sub-6 and mmWave band only.} We consider DeepMIMO's  I3 scenario \cite{deepMIMo_ahmed_dataset_gen_paper} for indoor LOS and NLOS. Precisely, it is an indoor conference room scenario with its hallways. It has two access points inside the conference room at 2 m height. For indoor, we consider operating frequencies 2.4 GHz and 60 GHz. For LOS, we consider R1 to R551, while for NLOS, we consider R552 to R1159 and BS1 (usually referred as access point for indoor) for both LOS and NLOS.  We consider 20 users moving along a similar trajectory at human walk speed of 1.3 m/s. The spacing between the users was 0.014 for LOS and 0.01 for NLOS. Out of 20 users, we consider the trajectory of the 11th user for testing. We would like to highlight that since the LOS and NLOS grids are completely different physical space, we did not implement the blockage segment for this particular setting, rather we just consider LOS and NLOS as two different extreme case scenarios.} 

\textcolor{black}{We further validate the proposed algorithm in indoor settings. 
%Due to dataset availability, our study is restricted to sub-6 GHz and mmWave bands. 
We use the I3 scenario from DeepMIMO \cite{deepMIMo_ahmed_dataset_gen_paper}, which represents an indoor conference room with adjoining hallways. The setup includes two AP's placed at a height of 2 m. We consider carrier frequencies of 2.4 GHz and 60 GHz. For LOS, we consider rows R1–R551, while for NLOS we consider R552–R1159, with AP1 serving both cases. A total of 20 UEs move along a common trajectory at a walking speed of 1.3 m/s. The inter-user spacing is 0.014 m for LOS and 0.01 m for NLOS. Out of 20 UEs, we consider the trajectory of the 11th UE for testing. Since the LOS and NLOS grids represent distinct physical environments, we do not apply the blockage-segment model here; instead, we treat LOS and NLOS as two separate extreme-case scenarios.}

%The methodology to construct dataset is exactly the same as DeepMIMO outdoor scenario in Sec. IV-A(1) and we skip for the sake of brevity.
\subsubsection{NYUSIM InH dataset (LOS and NLOS)} %\textcolor{black}{In case of NYUSIM, we consider indoor hotspot for offices (InH) as appropriate indoor scenario. The motivation to use NYC inh scenario is that that it has the capability of testing on or using sub 6 gigahertz, millimeter wave, and sub terahertz frequencies indoor. We use 10 users trajectories with each user separated by 0.01 m. We use the user 6th trajectory for testing. Given that this data set has the capability of generating of having the same physical scenario our space what's this,, we apply the blockage segment algorithm to generate low medium and high blockage scenarios for NYC and state as well.}
\textcolor{black}{For NYUSIM, we consider the Indoor Hotspot (InH) office scenario as the representative indoor case. The motivation for using InH is its ability to support evaluation at sub-6 GHz, mmWave, and sub-THz frequencies in indoor environments. We generate trajectories for 10 UEs, each separated by 0.01 m, and select the trajectory of the 6th UE for testing. Although this dataset allows LOS and NLOS realizations within the same physical layout, to maintain consistency with the DeepMIMO I3 scenario, we only consider LOS and NLOS as two scenarios for InH settings.}

\subsection{Empirical multiband testbed:}
%\textcolor{black}{To further validate the proposed scheme, we developed a multiband empirical testbed setup at the wireless and signal processing (WiSP) lab at Boston College, as shown in Figure \ref{multiband_setup}. Multiband transmitter is shown on the right while the receiver is on the left. This setup is capable of transmitting indoor at sub-6 GHz and at 60 GHz mmWave frequency. At the TX side, a PC running GNURadio companion is connected to USRP B-210 (UBX-160 daughter card) through USB for transmitting at sub-6 GHz. PC is also connected to USRP  N-210 (LFTX daughter card) through ethernet cable for transmitting at the mmWave Band. \textcolor{red}{Add 1 statement abt the baseband signal generated at GNURadio.} For sub-6 GHz, 2.4 GHz band is used while for the mmWave band the transmission is at 60 GHz. We read the same baseband signal from GNU Radio Companion, pass it through signal to differential amplifier, up-convert it to 60 GHz, and then transmitted over the air through horn antenna. At the receiving end, another PC running GNURadio is connected to USRP X-310 which has two daughter cards mounted. The received signal is received through sub-6 antenna and Horn antenna respectively. For the signal that is received through Horn antenna, it is down converted, passed through differential to single ended amplifier and then fed to LFRX. We would like to highlight that the receiver is set up on a movable cart. We move the cart back and forth at a human walking speed and collected few minutes of walking data. }

\textcolor{black}{To experimentally validate the proposed scheme, we developed a multiband empirical testbed setup at the Wireless and Signal Processing (WiSP) Laboratory, Boston College, as shown in Fig. \ref{multiband_setup}. The transmitter and receiver units are positioned on the right and left sides of the setup, respectively. The system supports indoor wireless transmission at both sub-6 GHz and 60 GHz mmWave bands.
On the transmitter side, a host PC running GNU Radio interfaces with a USRP B210 
%equipped with a UBX-160 daughterboard 
via USB for sub-6 GHz transmission. For mmWave transmission, the same PC connects to a USRP N210 with an LFTX daughterboard over Ethernet. The baseband signal is synthesized within GNU Radio using a custom flowgraph comprising modulation, filtering, and framing blocks.  For sub-6 GHz, we use 2.4 GHz ISM band, while the mmWave transmission is centered at 60 GHz. The baseband signal is routed through a single-ended differential amplifier, upconverted to 60 GHz, and transmitted via a horn antenna.}

\textcolor{black}{At the multiband receiver (on the left side of Fig. \ref{multiband_setup}), a second PC running GNU Radio interfaces with a USRP X310, which hosts two daughterboards,  the LFRX and UBX 160. The received signals are captured using a sub-6 GHz antenna and a isotropic antenna for mmWave reception. The mmWave signal is downconverted, passed through a differential-to-single-ended amplifier, and subsequently fed into the LFRX input of the USRP, whereas the sub-6GHz signal from the antenna is connected to the UBX 160 daughtercard input.
To emulate realistic mobility conditions, the receiver platform is mounted on a movable cart. The cart is manually moved back and forth at typical human walking speeds, and several minutes of multiband channel data were collected during this motion.}

\textcolor{black}{Furthermore, since data acquisition at sub-THz frequencies was infeasible due to unavailability of sub-THz hardware, we generated an emulated sub-Thz dataset for analysis. The emulation assumes that sub-Thz propagation exhibits characteristics similar to mmWave. Accordingly, existing mmWave channel data were adopted as the reference, and key parameters such as bandwidth and noise power were adjusted to approximate sub-THz conditions. Although this approximation does not fully capture the physical behavior of sub-THz propagation, it provides a reasonable \textcolor{black}{lower} bound on power consumption for three band indoor settings.}

%Since the BS  has $32$ antennas, $32$ complex channel coefficients are obtained at every time instant. Thus, the channel coefficient matrix for each outdoor dataset scenario was $2751 \times 32$. Based on channel gains, the SNR and hence the rate is computed as per $(\ref{snr_eqn})$ and $(\ref{rate_eqn})$ respectively, and its shape would be $2751 \times 1$.

%\vspace{-2mm}
% \usepackage{multirow}
% \usepackage{graphicx}

\subsection{Forecasting Methodology}
%\subsubsection{LSTM based forecasting}
% \subsubsection{Transformer based forecasting}
The UE location and channel are usually correlated with its previous history. This is also justifiable in real-life highway, urban road scenarios, \textcolor{black}{and in the indoor settings where the speed of UE is very low. 
%To confirm this, we plotted the autocorrelation of the location data points of UE with its past (lag) samples as shown in Fig. \ref{autocorrelation_plot}. As intuitive, the autocorrelation was found significantly high. 
%(This would also lead to $\ldots$-add one sentence regarding correlation in context of wireless context-take sir's help). 
This motivates us to utilize the LSTM as a forecaster.
Once the datasets are formulated, temporal sequences are constructed from the available channel data. The number of UEs considered in each scenario is  specified in the preceding subsections. 
%An important feature of the LSTM model is its ability to learn from history by looking back from the current time step. 
LSTM models learn from historical context.
In our experiments, hyperparameter tuning led us to select a lookback of 15. Thus, the LSTM uses the past 15 samples to forecast either the rate or $|\text{\textbf{h}}_i|^2$ up to $T-1$ time slots into the future. We would like to highlight that we use direct multistep ahead forecasting that forecasts all the $T-1$ steps directly instead of iterative multistep ahead, since in the latter, the error accumulates, resulting in poor predictions \cite{transforemr_for_time_series_AAAI23}.}
\begin{table}[t!]
\centering
\caption{Stacked LSTM Hyperparameters }
\renewcommand{\arraystretch}{1}
\begin{tabular}{|c|c|}
\hline
\textbf{Hyperparameter} &  \textbf{Value} \\ \hline
%Initial learning rate ($\alpha$) & 0.0001 \\ \hline
learning rate; Batch size; Epochs & 0.0001 ; 64 ;  50 \\ \hline
%Number of epochs & 50 \\ \hline
Number of layers (Depth) & 4 hidden layers + 1 dense layer \\ \hline
Number of LSTM units (Width) & \{100, 64, 64, 32\} \\ \hline
Dropout & 0.4 between hidden  layers   \\ \hline
Optimizer ; Loss function & Adam ; Mean square error (MSE) \\ \hline
%Loss function & Mean square error (MSE) \\ \hline
%Activation function & \begin{tabular}[c]{@{}c@{}}ReLU (for hidden layer); \\ Sigmoid (for o/p layer)\end{tabular} \\ \hline
Activation function &  ReLU (for hidden layers) \\ \hline
\end{tabular}
\label{lstm_hyperparameter_table}
\end{table}
\begin{table}[t!]
\centering
\caption{Transformer Hyperparameters}
\renewcommand{\arraystretch}{1}
\begin{tabular}{|c|c|}
\hline
\textbf{\textbf{Hyperparameter}} & \textbf{\textbf{Value}} \\ \hline
Batch size; Epochs               & 256 ;  10               \\ \hline
\# of Encoder \& Decoder layers  & 4 layers (each)         \\ \hline
 Dimensionality of the model (d\_model) & 32       \\ \hline
% \begin{tabular}[c]{@{}c@{}}Dimensionality of the model  \\ and its embeddings (d\_model)\end{tabular} & 32 
%\\ \hline
Optimizer ; learning rate ; $\beta_1$ ;  $\beta_2$      & AdamW ; 0.0006  ; 0.9 ; 0.95        \\ \hline
\end{tabular}
\label{transformer_hyperparameter_table}
\end{table}

\begin{table*}[t!]
\centering
\caption{Summary of the forecasting errors for the test sequences on DeepMIMO - O1. Lower the value, better is the prediction. }
%\resizebox{\textwidth}{!}{%
\renewcommand{\arraystretch}{1.15}
\begin{tabular}{c|c|c|cccc|cccc}
\hline
 & \textbf{}                           & \textbf{Metrics}     & \multicolumn{4}{c|}{\textbf{NRMSE}}                                  & \multicolumn{4}{c}{\textbf{NMAE}} \\ \cline{2-11} 
\multirow{-2}{*}{\textbf{Approach}} &
  \textbf{Bands} &
  \textbf{Slots} &
  \textbf{Step 4} &
  \textbf{Step 3} &
  \textbf{Step 2} &
  \textbf{Step 1} &
  \textbf{Step 4} &
  \textbf{Step 3} &
  \textbf{Step 2} &
  \textbf{Step 1} \\ \hline
 &                                     & \textbf{LSTM}        & 0.0620 &	0.0598 &	0.0541 &	0.0494  & 0.0428 &	0.0388 &	0.0353 &	0.0309   \\ \cline{3-11} 
 & \multirow{-2}{*}{\textbf{Sub-6 GHz}} & \textbf{Transformer}& 0.0615	& 0.0527	& 0.0305	& 0.0181  & 0.0439	& 0.0398 &	0.0231 &	0.0145           \\ \cline{2-11} 
 &                                     & \textbf{LSTM}        & 0.1765	& 0.1717 &	0.1698	& 0.1603  & 0.1366  &	0.1334  &	0.1332  &	0.1247   \\ \cline{3-11} 
 & \multirow{-2}{*}{\textbf{mmWave}}   & \textbf{Transformer} & 0.2269 &	0.1984 &	0.1872 &	0.1354  & 0.1723 &	0.1515 &	0.1462 &	0.1005            \\ \cline{2-11} 
 &                                     & \textbf{LSTM}    & 0.3089 &	0.3035	& 0.3020 &	0.2972  & 0.2297 &	0.2225 &	0.2164 &	0.2015   \\ \cline{3-11} 
\multirow{-6}{*}{\textbf{I: Rate forecaster}} &
  \multirow{-2}{*}{\textbf{THz}} &
  \textbf{Transformer} &
  0.4263 &
  0.3908 &
  0.3386 &
  0.2200 & 0.3417 &	0.3144 &	0.2657 &	0.1656
   \\ \hline
 &                                     & \textbf{LSTM} & 0.4948	&0.4623&	0.3832&	0.2483	&0.2405	&0.2084	&0.1672	&0.1062   \\ \cline{3-11} 
 & \multirow{-2}{*}{\textbf{Sub-6 GHz}} & \textbf{Transformer} & 0.6826&	0.5870&	0.3798&	0.1757&	0.3405&	0.2950&	0.1783&	0.0805  \\ \cline{2-11} 
 &                                     & \textbf{LSTM}        & 1.2307&	1.1419&	1.0802&	1.0222&	0.5460&	0.5440&	0.5420&	0.5358     \\ \cline{3-11} 
 & \multirow{-2}{*}{\textbf{mmWave}}   & \textbf{Transformer} & 1.2126&	1.0751&	0.8754&	0.6849&	0.5873&	0.5208&	0.4264&	0.3297    \\ \cline{2-11} 
 &                                     & \textbf{LSTM}        & 0.7763&	0.7702	&0.7670&	0.6842&	0.3974&	0.3843&	0.3774&	0.3767     \\ \cline{3-11} 
\multirow{-6}{*}{\textbf{II: Channel forecaster}} &
  \multirow{-2}{*}{\textbf{THz}} &
  \textbf{Transformer} &
  0.8242	&0.8224&	0.7758&	0.5490&	0.4625&	0.4509&	0.4048&	0.2959
   \\ \hline
\end{tabular}
\label{table_summary_of_forecast_errors}
\end{table*}

\begin{table*}[t!]
\centering
\caption{\textcolor{black}{Summary of the forecasting errors for the test sequences on DeepMIMO-I3}. Lower the value, better is the prediction.}
\renewcommand{\arraystretch}{1.15}
\begin{tabular}{c|c|c|cccc|cccc|c}
\hline
 & \textbf{} & \textbf{Metrics} & \multicolumn{4}{c|}{\textbf{NRMSE}} & \multicolumn{4}{c|}{\textbf{NMAE}} & \\ \cline{2-12} 
\multirow{-2}{*}{\textbf{Approach}} &
  \textbf{Bands} &
  \textbf{Slots} &
  \textbf{Step 4} &
  \textbf{Step 3} &
  \textbf{Step 2} &
  \textbf{Step 1} &
  \textbf{Step 4} &
  \textbf{Step 3} &
  \textbf{Step 2} &
  \textbf{Step 1} &
  \textbf{LOS/NLOS} \\ \hline
\multirow{8}{*}{\textbf{I: Rate forecaster}} & \multirow{4}{*}{\textbf{Sub-6 GHz}} & \multirow{2}{*}{\textbf{LSTM}} & 0.0174 & 0.0152 & 0.0130 & 0.0109 & 0.0100 & 0.0080 & 0.0064 & 0.0056 & LOS \\ \cline{4-12}
 & & & 0.0183 & 0.0156 & 0.0141 & 0.0108 & 0.0099 & 0.0083 & 0.0073 & 0.0047 & NLOS \\ \cline{3-12}
 & & \multirow{2}{*}{\textbf{Transformer}} & 0.0184 & 0.0162 & 0.0144 & 0.0115 & 0.0115 & 0.0095 & 0.0081 & 0.0060 & LOS \\ \cline{4-12}
 & & & 0.0178 & 0.0163 & 0.0155 & 0.0124 & 0.0112 & 0.0099 & 0.0102 & 0.0069 & NLOS \\ \cline{2-12}
 & \multirow{4}{*}{\textbf{mmWave}} & \multirow{2}{*}{\textbf{LSTM}} & 0.0158 & 0.0187 & 0.0170 & 0.0146 & 0.0099 & 0.0103 & 0.0087 & 0.0077 & LOS \\ \cline{4-12}
 & & & 0.0347 & 0.0307 & 0.0267 & 0.0213 & 0.0269 & 0.0233 & 0.0200 & 0.0152 & NLOS \\ \cline{3-12}
 & & \multirow{2}{*}{\textbf{Transformer}} & 0.0183 & 0.0170 & 0.0150 & 0.0118 & 0.0099 & 0.0086 & 0.0071 & 0.0051 & LOS \\ \cline{4-12}
 & & & 0.0714 & 0.0582 & 0.0517 & 0.0384 & 0.0573 & 0.0451 & 0.0398 & 0.0288 & NLOS \\ \hline
\multirow{8}{*}{\textbf{II: Channel forecaster}} & \multirow{4}{*}{\textbf{Sub-6 GHz}} & \multirow{2}{*}{\textbf{LSTM}} & 0.1742 & 0.1703 & 0.1393 & 0.1093 & 0.0957 & 0.0937 & 0.0647 & 0.0465 & LOS \\ \cline{4-12}
 & & & 0.1261 & 0.1319 & 0.1020 & 0.0851 & 0.0744 & 0.0889 & 0.0598 & 0.0515 & NLOS \\ \cline{3-12}
 & & \multirow{2}{*}{\textbf{Transformer}} & 0.2001& 0.1752 & 0.1574 & 0.1224 & 0.1246 & 0.1009 & 0.0866 & 0.0658 & LOS \\ \cline{4-12}
 & & & 0.1582 & 0.1459 & 0.1556 & 0.0980 & 0.1033 & 0922 & 0.1065 & 0.0603 & NLOS \\ \cline{2-12}
 & \multirow{4}{*}{\textbf{mmWave}} & \multirow{2}{*}{\textbf{LSTM}} & 0.0637 & 0.0623 & 0.0507 & 0.0401 & 0.0328 & 0.0343 & 0.0252 & 0.0176 & LOS \\ \cline{4-12}
 & & & 0.1401 & 0.1413 & 0.1074 & 0.0917 & 0.0987 & 0.1043 & 0.0789 & 0.0673 & NLOS \\ \cline{3-12}
 & & \multirow{2}{*}{\textbf{Transformer}} & 0.0813 & 0.0756 & 0.0688 & 0.0569 & 0.0456 & 0.0397 & 0.0382 & 0.0257 & LOS \\ \cline{4-12}
 & & & 0.3124 & 0.2590 & 0.2250 & 0.1517 & 0.2450 & 0.1936 & 0.1685 & 0.1161 & NLOS \\ \hline
\end{tabular}
\label{Table:summaryOfDeepMIMOIndoorPreds}
\end{table*}

\textcolor{black}{The specific numbers of training, validation, and test sequences for each dataset are summarized in Table~\ref{summary_of_sequences}. Since the availability of datapoints varies significantly across different datasets, we ensured that the number of sequences used across scenarios remains reasonably consistent. We skip a few intermediate sequences between training and validation to avoid data leakage. This approach is consistent with a realistic system setting, where the BS predicts future channel behavior based on its recent history and the characteristics of users in its vicinity.}

We built a stacked LSTM forecaster(s), that essentially learns a mapping function between past and future rates, and provides  forecasts of the rates and $|\text{\textbf{h}}_i|^2$ to help achieve the objective in \eqref{eq:power_opt_a}. The LSTM hyperparameters are summarized in Table \ref{lstm_hyperparameter_table}.
In addition to the LSTM based predictor (which performs the best as shown subsequently), we considered a simpler predictor we call \textbf{current channel prediction}  and Transformer based forecast.
For the current channel prediction, the rate/channel in the current slot is used as the forecast for the remaining slots in the frame. E.g., for the rate predictor, in slot $t$ of a given frame $\hat{R}_i[t+1,t] = \cdots = \hat{R}_i[T, t] = R_i[t]$, where $\hat{R}_i[\ell,t]$ is the predicted rate of slot $\ell$ at band $i$  with  $t$ denoting the slot when the prediction was made. The Transformer based predictor is based on \cite{vaswani2017attention}, which has recently received significant attention in the literature. The Transformer hyperparameters are summarized in Table \ref{transformer_hyperparameter_table}.
\section{Numerical Results}
\label{secV}
We use  `DeepMIMO' \cite{deepMIMo_ahmed_dataset_gen_paper} and `NYUSIM' datasets \cite{nyusim4} to generate channels, Keras API \cite{keras} with TensorFlow backend to create the stacked LSTM as a forecaster, while Transformer was built using Pytorch.
%As described earlier in Fig. \ref{autocorrelation_plot},  Non zero autocorrelation demonstrates that the data samples are temporally correlated. This is because the autocorrelation value does not reduce to zero instantly. This temporal correlation is exploited in this work using LSTM based sensing framework.
%The list of simulation parameters is summarized in Table \ref{deepmimo_parameters}. 
%We consider the beam training time $T_{beam}=1 \mu s$, $\alpha \approx \text{Uniform}(0,\pi),$ $\theta:=102/M_y,$ $N_{CB}:= M_y,$ $\rho = 0.6,$ 
We use UE noise figure $= 7$ dB, $\sigma_i^2=$K$\cdot$Temp$\cdot$$B_i$$\cdot$UE noise figure, where K is Boltzmann's constant, and Temp. $= 300$  Kelvin. Switching cost $\nu = 0.05$. The power consumed in each band is as per Section II-B.
%As mentioned earlier, the objective in eqn. (\ref{M_based_eqn}) can be fulfilled using rate forecaster and channel forecaster. 
To better quantify prediction errors, we use normalized root mean square error (NRMSE) and normalized mean absolute error (NMAE) as forecaster metrics.
We compare the proposed stacked LSTM forecaster with the Transformer model, which uses the attention mechanism to capture context. The hyperparameters were tuned for fair comparison and the lookback size of 15 was used for each sequence. 

Table \ref{table_summary_of_forecast_errors} summarizes the forecasting errors for both the approaches for DeepMIMO outdoor scenario.  We can notice that predictions using the rate forecaster is comparatively more accurate than the channel forecaster. This is intuitive, and due to the fact that the fluctuations in the channel   are significantly higher as compared to the rate fluctuations. Furthermore, we can notice that LSTM predictions are much accurate than those of the Transformer, following similar observation in \cite{transforemr_for_time_series_AAAI23}. \textcolor{black}{For instance, across both forecasting tasks, LSTM achieves consistently lower  prediction error than Transformer for all bands. For rate forecasting with Step-4 ahead prediction, LSTM improves performance by $\sim$30\% in NRMSE and $\sim$27\% in NMAE on an average across Sub-6, mmWave, and THz. Similarly, for channel forecasting, LSTM has $\sim$20\% lower NRMSE and $\sim$15\% lower NMAE.}
% \textcolor{black}{In the following, the results for the current channel prediction approach only includes the forecasts using LSTM.} 

%\textcolor{black}{Table \ref{Table:summaryOfDeepMIMOIndoorPreds} summarizes the forecasting errors for DeepMIMO Indoor scenario. For Step 4 predictions across Sub6 GHz and mmWave bands, LSTM consistently outperforms Transformer in both Rate and Channel forecasters. In the rate forecaster, LSTM achieves 0.01785 in Sub6 GHz and 0.02525 in mmWave, compared to 0.0181 and 0.04485 for Transformer. The Channel forecaster shows higher errors, with LSTM at 0.15015 (Sub6 GHz) and 0.1019 (mmWave) versus 0.17915 and 0.19685 for Transformer. Moreover, as intuitive, performance is consistently better under LOS than NLOS; for example, mmWave Rate Step 4 NRMSE for LSTM increases from 0.0158 (LOS) to 0.0347 (NLOS).}
\textcolor{black}{Table \ref{Table:summaryOfDeepMIMOIndoorPreds} summarizes the forecasting performance for the DeepMIMO indoor scenario. For Step 4 predictions across Sub-6 GHz and mmWave bands, LSTM consistently outperforms the Transformer for both rate and channel forecasters when considering average NRMSE across LOS and NLOS conditions. In rate forecasting, LSTM yields comparable performance in Sub-6 GHz with a modest 1–2\% reduction in error, while achieving a substantial $\sim$44\% average error reduction in mmWave. For channel forecasting, the gains are more pronounced, with LSTM reducing average errors by $\sim$16\% in Sub-6 GHz and $\sim$48\% in mmWave relative to the Transformer. As expected, performance is consistently better under LOS than NLOS; for example, in mmWave Rate forecasting, the Step 4 NRMSE increases by nearly 120\% when moving from LOS to NLOS.}
\begin{figure}[t!]
\centering
\includegraphics[scale=0.45]{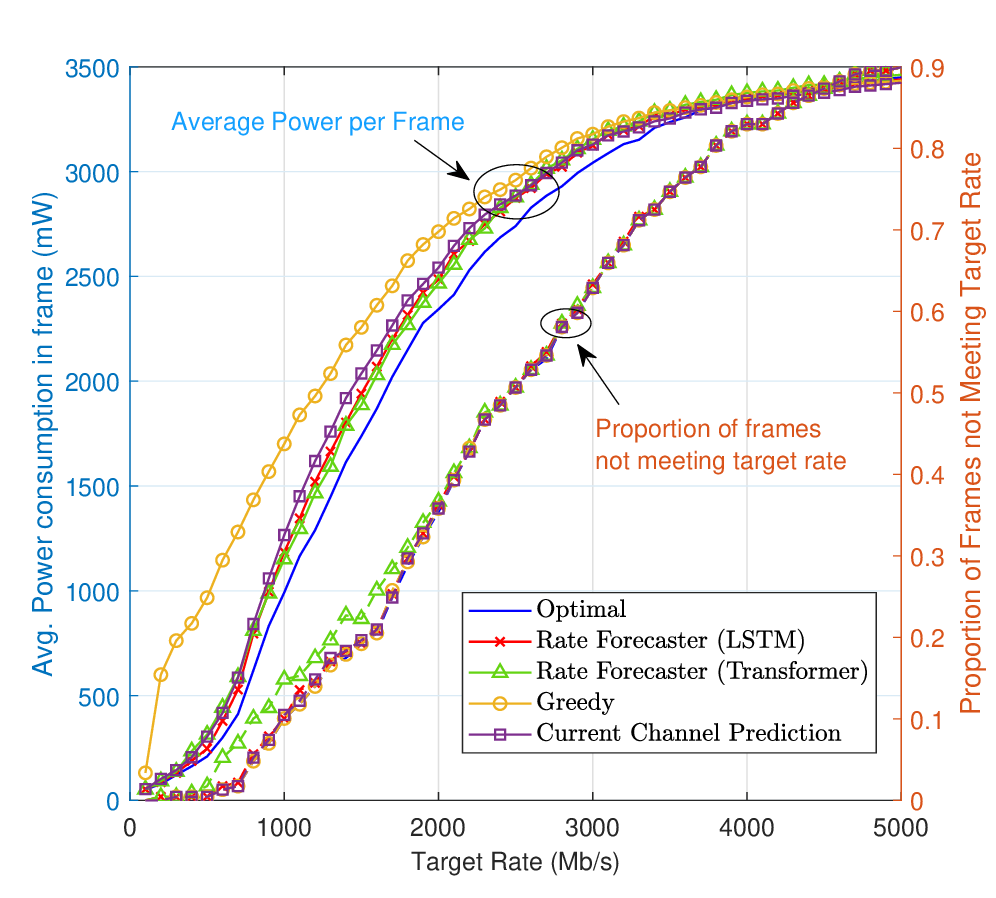}
\caption{\small{Average Power Consumed vs. Target Rate (Rate forecaster).}}
%The fraction of frames with rates below the target rate is also plotted against the right vertical axis. }}
\label{powerconsumed_vs_rate}
\end{figure}
\begin{figure}[t!]
\centering
\includegraphics[scale=0.48]
{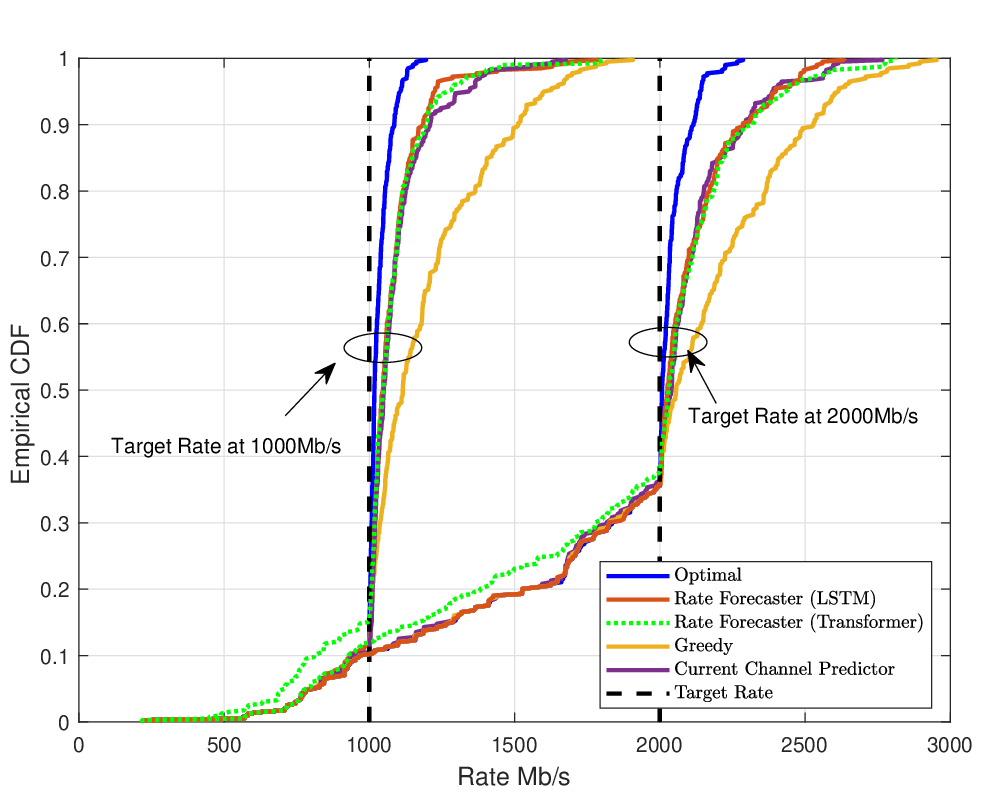}
\caption{\small{CDF of rates for 1000Mb/s; 2000Mb/s Target Rates.} \vspace{-5mm}}
\label{cdf_of_rates}
\end{figure}
\begin{figure}[t!]
\centering
\includegraphics[scale=0.44]{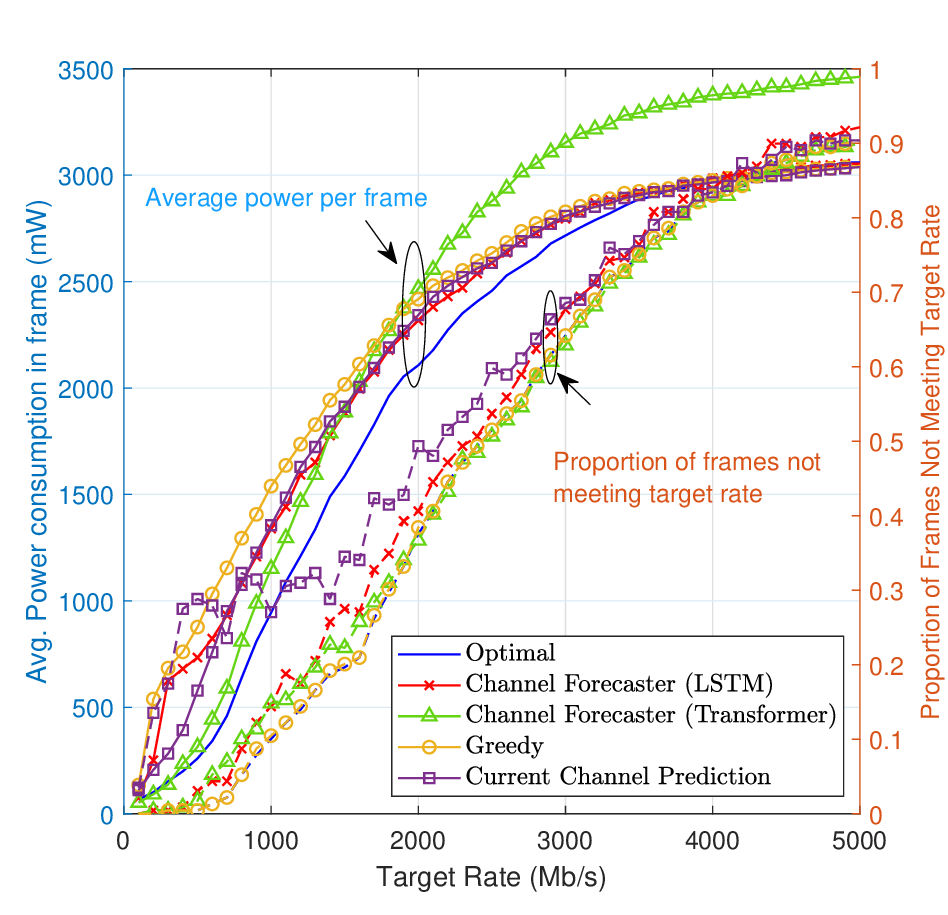}
\caption{\small{Avg. Power Consumed vs. Target Rate (Channel forecaster).}}
%. The fraction of frames with rates below the target rate is also plotted against the right vertical axis. }}
\label{powerconsumed_vs_rate_ChanEst}
\end{figure}

Although, to the best of our knowledge, there is no existing work with which we can fairly compare the proposed approach, we compare our approach with \textbf{i) Greedy approach:} In this scheme, which is the standard approach, at the start of each slot, the rates/channels in all bands are estimated. With a possible exception to be described subsequently, the BS picks the band with the highest rate for that slot. Then, it subtracts the achieved rate in this slot from the target sum rate for the frame in preparation for the next time slot.  The exception is if the target sum rate for the frame can be achieved in the current slot by transmitting in a  lower-power band. In this case, the BS  picks  the lowest UE power consumption band which can achieve the sum rate. This approach is named greedy in a sense that the BS always chooses the highest rate band, except when the sum rate in the frame can be satisfied in the current slot with a lower power band. 
\textbf{ii) Optimal approach:} In this approach, the rates for all slots are known non-causally, and an exhaustive search is done on every frame to find the band assignments which minimizes power consumption at the UE while meeting the target rate. We also refer this interchangeably as ground truth. 
\begin{figure*}[t!]
    \centering
    \captionsetup{format =hang}%
   % \vskip\baselineskip
    \begin{subfigure}[t]{0.32\textwidth}  
        \centering 
        \includegraphics[width=\textwidth]{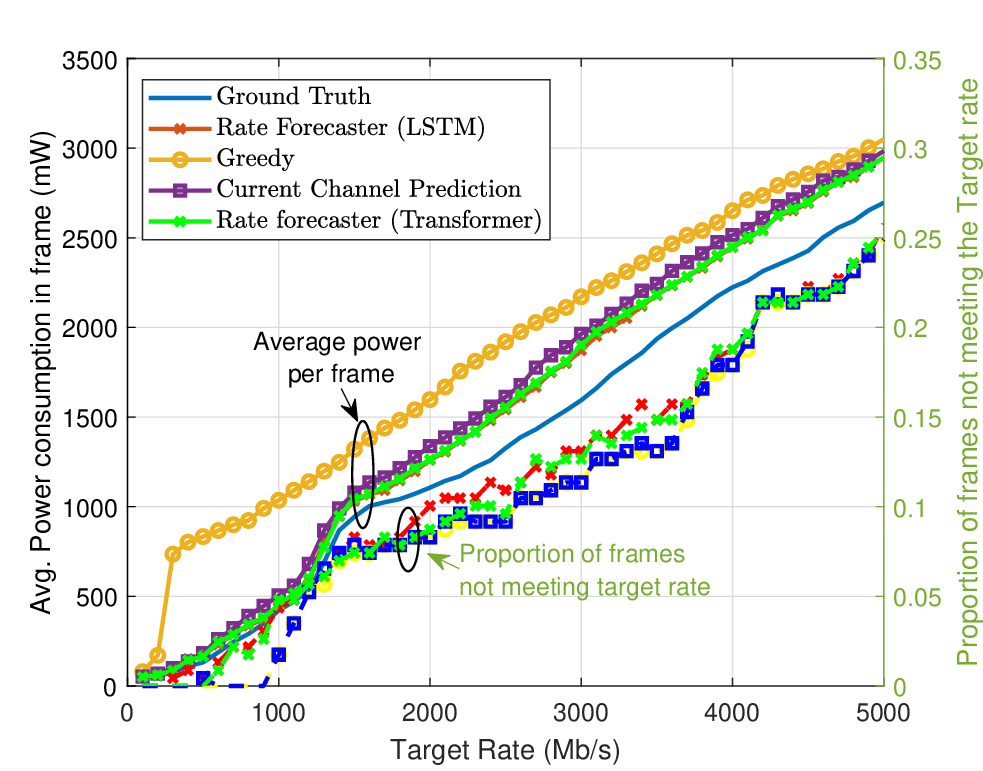}
        \caption[]%
        {{\small Low Blockage Scenario}}    
        \label{fig:BPSK_Pds}
    \end{subfigure}
%    \hspace{0.1cm}
%    \quad
    %\hfill
    \begin{subfigure}[t]{0.32\textwidth}  
        \centering 
        \includegraphics[width=\textwidth]{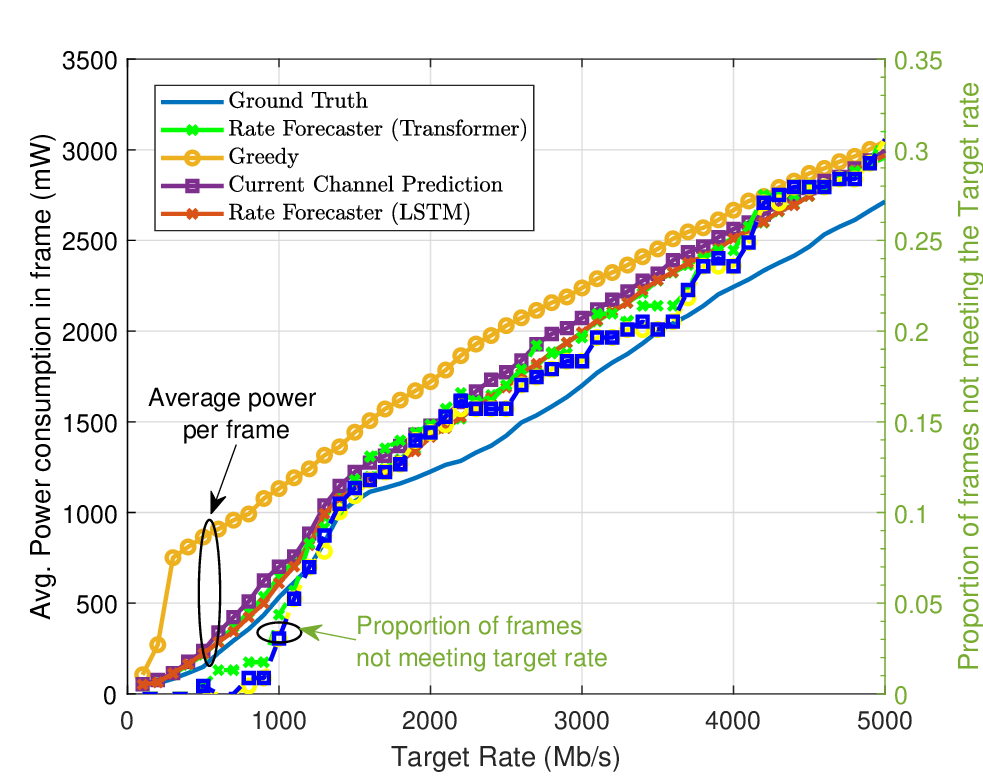}
        \caption[]%
        {{\small Medium Blockage scenario}}    
        \label{fig:QPSK_Pds}
    \end{subfigure}
%    \quad
    %\hfill
    \begin{subfigure}[t]{0.32\textwidth}  
        \centering 
        \includegraphics[width=\textwidth]{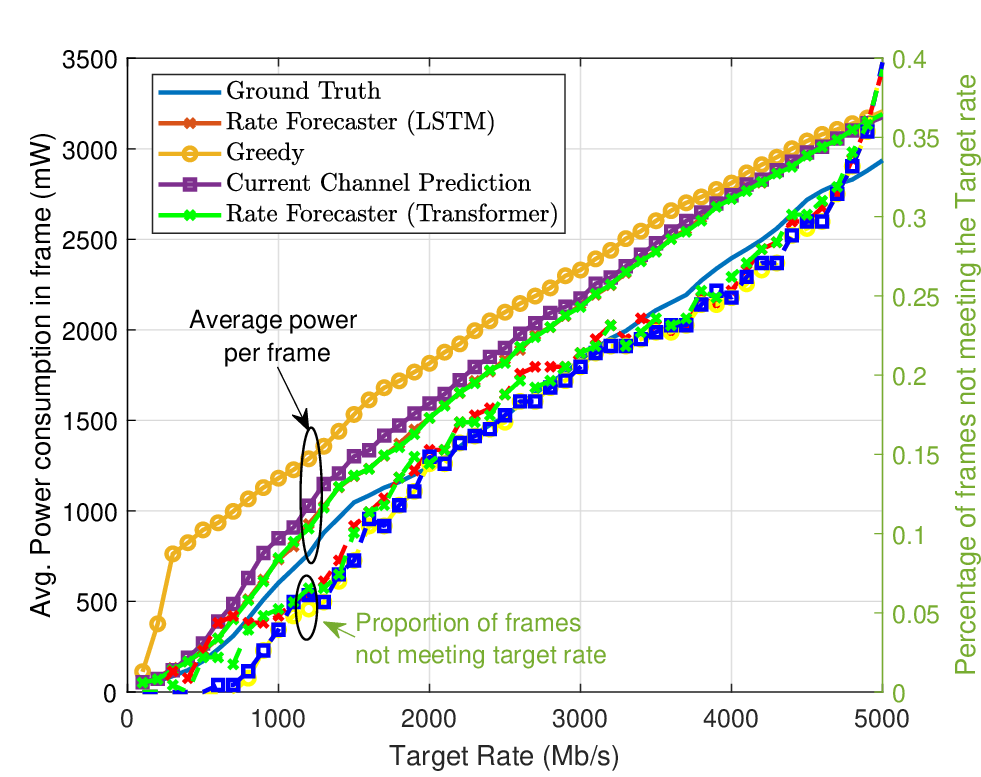}
        \caption[]%
        {{\small High Blockage Scenario}}    
        \label{fig:GFSK_Pds}
    \end{subfigure}
%    \quad
%    \hfill
    % \begin{subfigure}[t]{0.245\textwidth}  
    %     \centering 
    %     \includegraphics[width=\textwidth]{UHF_Pds.eps}
    %     \caption[]%
    %     {{\small UHF Band}}    
    %     \label{fig:QAM-64_Pds}
    % \end{subfigure}
%    \quad 
    \caption{\textcolor{black}{Average Power Consumed v/s Target Rate with rate forecaster for NYUSIM outdoor dataset with low, medium and high blockage scenarios.}}
    \label{nyusim_outdoor_lowmedhigh}
\end{figure*}

\begin{figure*}[t!]
    \centering
    \captionsetup{format =hang}%
   % \vskip\baselineskip
    \begin{subfigure}[t]{0.45\textwidth}  
        \centering 
        \includegraphics[width=\textwidth]{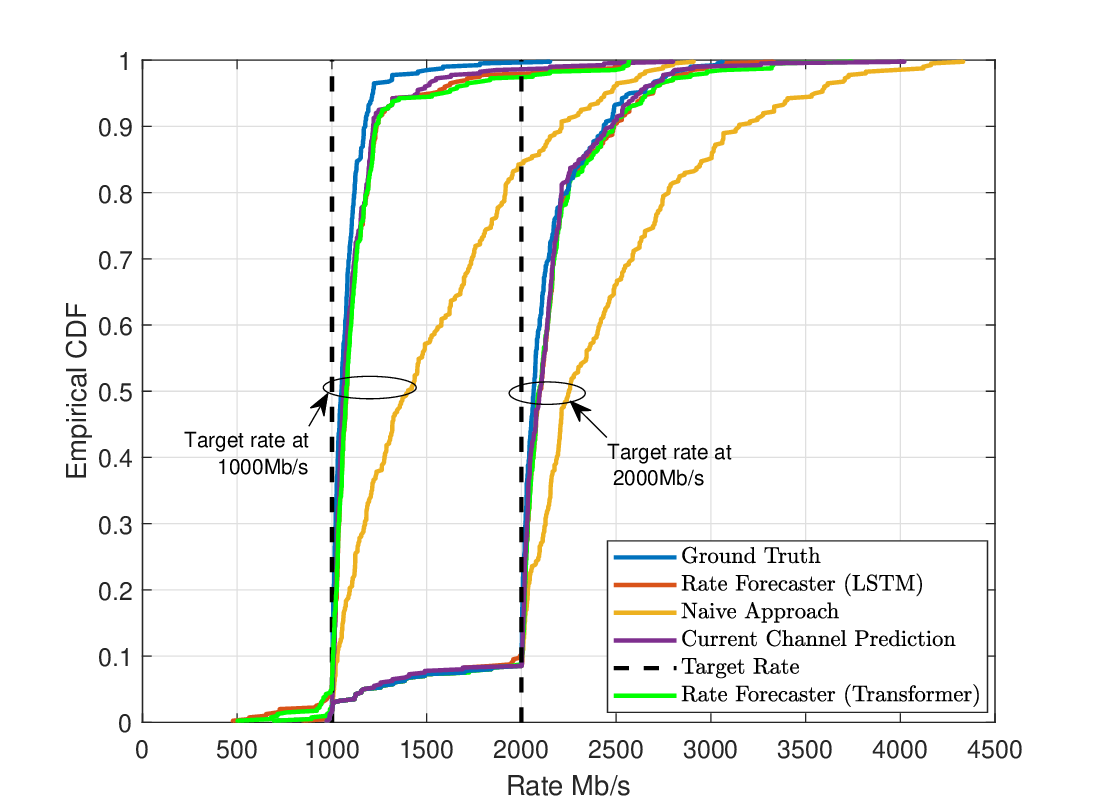}
        \caption[]%
        {{\small Low blockage scenario }}    
        \label{fig:BPSK_Pds}
    \end{subfigure}
%    \hspace{0.1cm}
%    \quad
    %\hfill
    \begin{subfigure}[t]{0.45\textwidth}  
        \centering 
        \includegraphics[width=\textwidth]{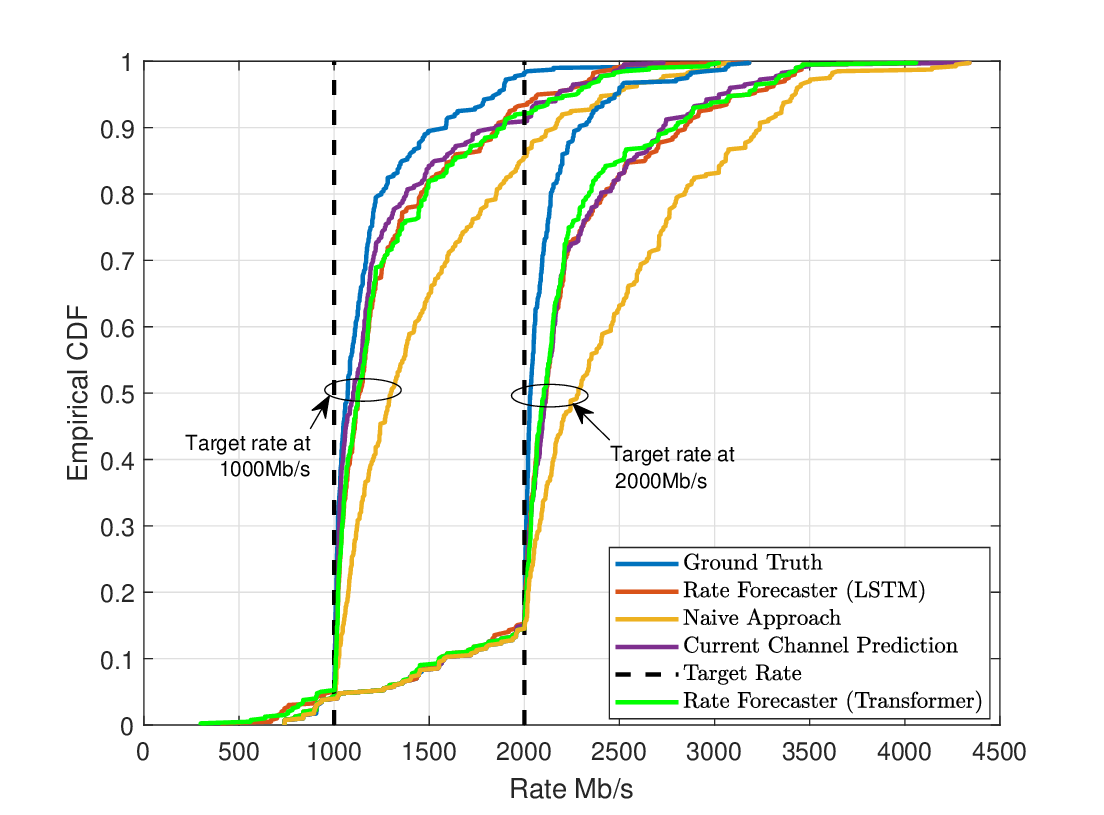}
        \caption[]%
        {{\small High blockage scenario}}    
        \label{fig:QPSK_Pds}
    \end{subfigure}
%    \quad 
    \caption{CDF of rates with rate forecaster for NYUSIM outdoor dataset under low and high blockage scenarios}
    \label{nyusim_cdf_low_highblockage_NYUSIMoutdoor}
\end{figure*}
Fig. \ref{powerconsumed_vs_rate}
shows the average power consumed per frame for various target rates, using different rate forecasters, greedy and optimal approaches.  The figure also shows the fraction of frames for which the target rate isn't met, with the axes labels on the right. Note that the proposed approach with LSTM tracks the optimal approach, and outperforms other considered approaches, except the transformer, in terms of average power consumption. Further, due to better rate predictions, the LSTM outperforms Transformer for almost all the target rates in terms of the fraction of frames meeting the target rate. Hence, even though the Transformer approach slightly outperforms the LSTM approach for the average power consumption at certain target rates, the significantly higher rate of frames not meeting target makes it less desirable than the LSTM approach. More specifically, with a target rate of 1500 Mb/s, 
%our approach with the LSTM rate forecaster consumes more than 300 mW less than the greedy band selection approach, and the current channel predictor consumes about 200 mW less than the greedy approach. 
our approach with the LSTM rate forecaster reduces power consumption by over 300 mW compared to the greedy band selection approach, while the current channel predictor achieves a reduction of $\sim$200 mW.
Further, the LSTM based approach is within 120 mW of the optimal, non-causal approach. Both the LSTM and the optimal approach have approximately the same proportion of frames not meeting the target rate of 1500 Mb/s. \textcolor{black}{E.g. for the LSTM and Optimal approach, the proportion of frames not meeting the target rate is $\sim$20\%, whereas for the transformer, it is $\sim$25\%.}  Since the proportion of frames which did not meet the target rate are approximately equal for all schemes except the Transformer, the savings in power from the LSTM and current-channel predictor do not come at the expense of a significant statistical reduction in data rates. Furthermore, to help quantify the distribution of rates actually achieved, we plot the CDFs of the rates with the rate forecaster in Fig. \ref{cdf_of_rates} with target rates as 1000 Mb/s and 2000 Mb/s. The key observation here is that the distribution of rates below the threshold is approximately equal for all approaches except  with the Transformer-based rate forecaster, which has an appreciably worse CDF below the target rates. 
\begin{figure*}[t!]
    \centering
    \captionsetup{format =hang}%
   % \vskip\baselineskip
    \begin{subfigure}[t]{0.425\textwidth}  
        \centering 
        \includegraphics[width=\textwidth]{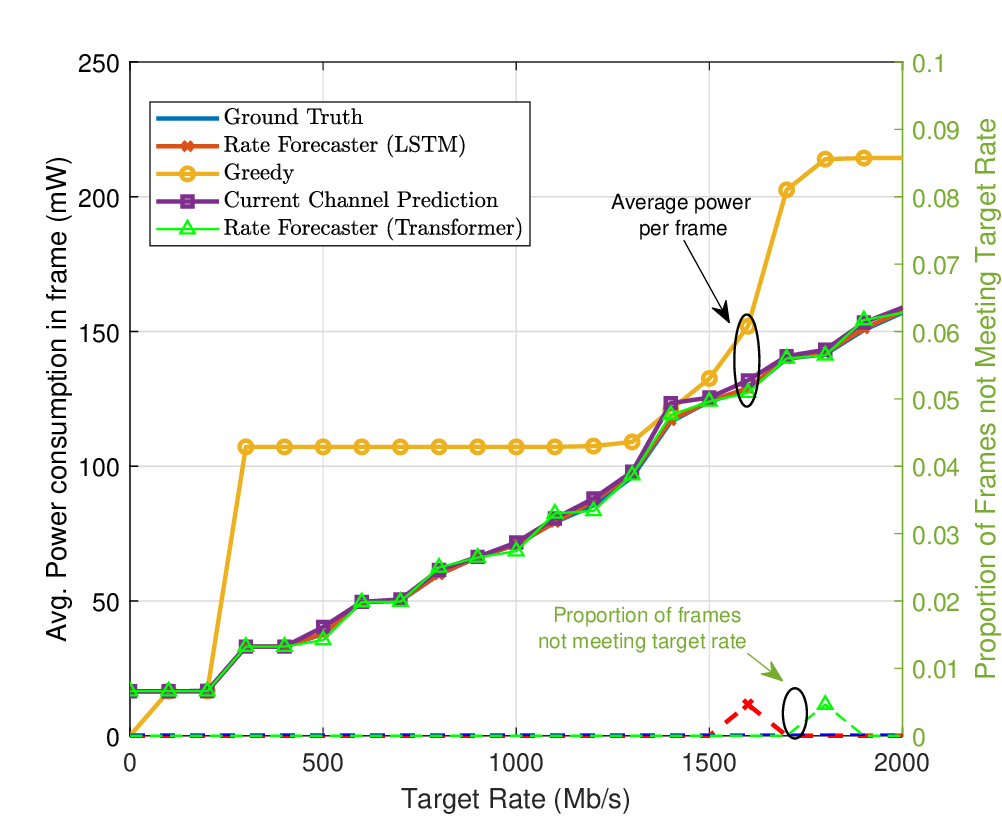}
        \caption[]%
        {{\small LOS scenario }}    
        \label{fig:BPSK_Pds}
    \end{subfigure}
%    \hspace{0.1cm}
%    \quad
    %\hfill
    \begin{subfigure}[t]{0.425\textwidth}  
        \centering 
        \includegraphics[width=\textwidth]{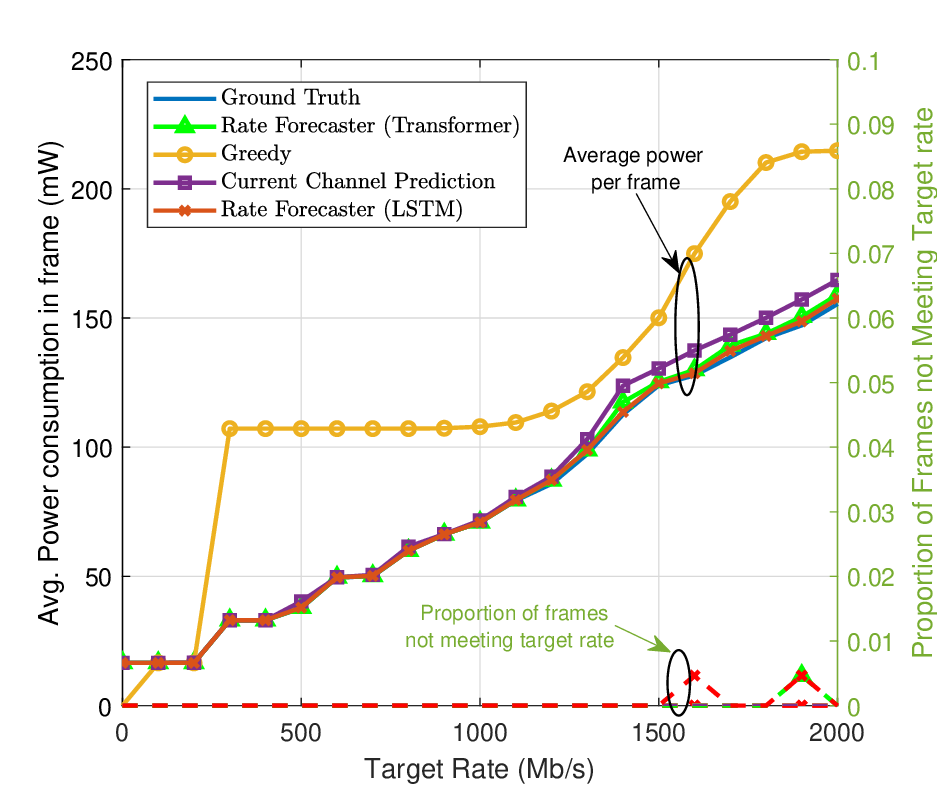}
        \caption[]%
        {{\small NLOS scenario}}    
        \label{fig:QPSK_Pds}
    \end{subfigure}
%    \quad 
    \caption{Average Power Consumed v/s Target Rate with rate forecaster. DeepMIMO I3 Indoor dataset with LOS and NLOS}
    \label{deepmimo_rateforecaster_LOSandNLOSIndoor}
\end{figure*}

\begin{figure*}[t!]
    \centering
    \captionsetup{format =hang}%
   % \vskip\baselineskip
    \begin{subfigure}[t]{0.32\textwidth}  
        \centering 
        \includegraphics[width=\textwidth]{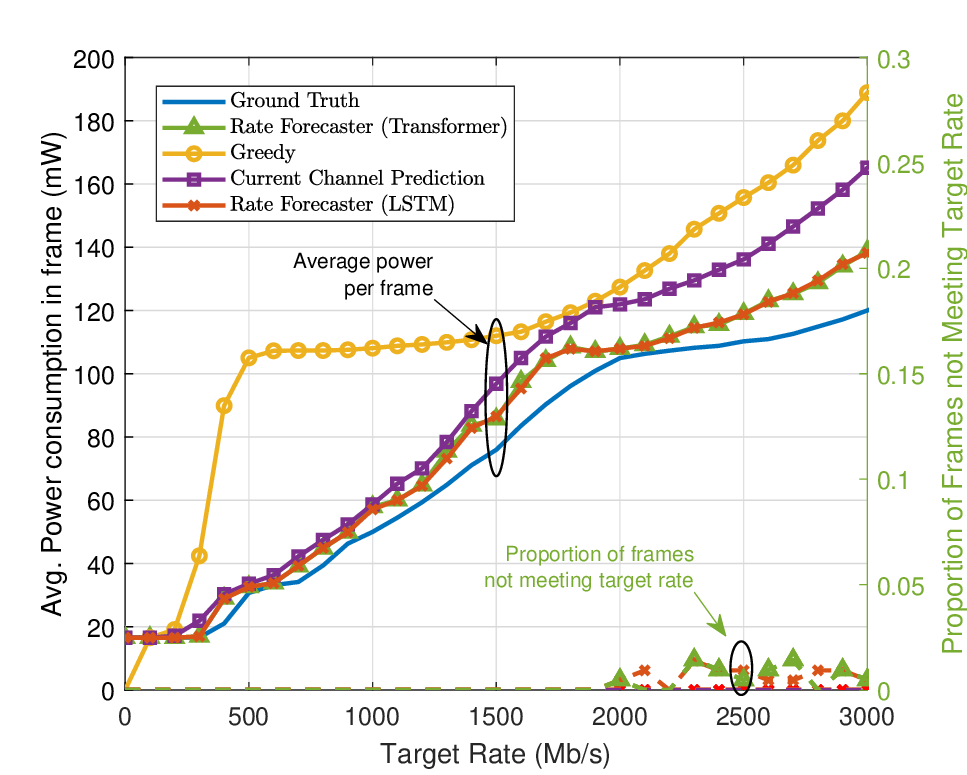}
        \caption[]%
        {{\small LOS Scenario - 2 bands}}    
        \label{fig:BPSK_Pds}
    \end{subfigure}
%    \hspace{0.1cm}
%    \quad
    %\hfill
    \begin{subfigure}[t]{0.32\textwidth}  
        \centering 
        \includegraphics[width=\textwidth]{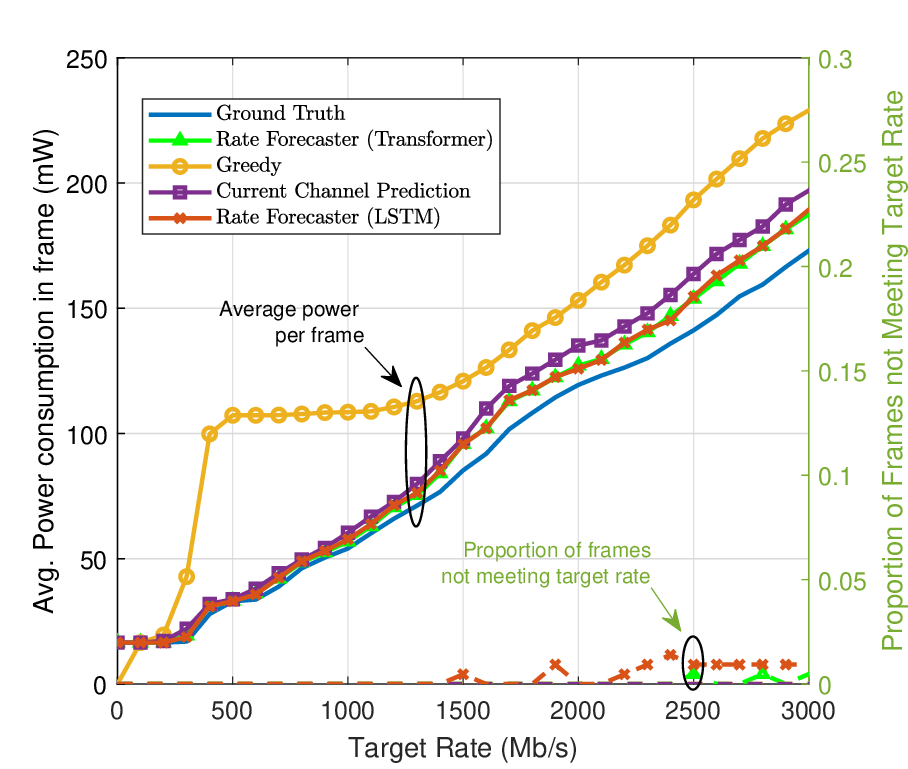}
        \caption[]%
        {{\small NLOS scenario - 2 bands}}    
        \label{fig:QPSK_Pds}
    \end{subfigure}
%    \quad
    %\hfill
    \begin{subfigure}[t]{0.33\textwidth}  
        \centering 
        \includegraphics[width=\textwidth]{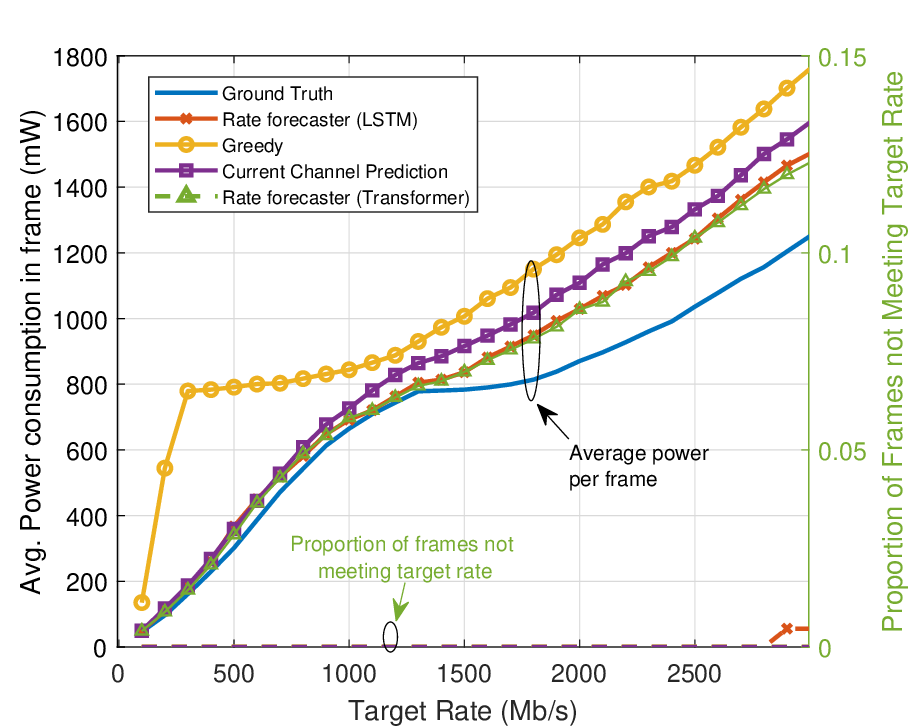}
        \caption[]%
        {{\small NLOS Scenario - 3 bands}}    
        \label{fig:GFSK_Pds}
    \end{subfigure}
%    \quad
%    \hfill
    % \begin{subfigure}[t]{0.245\textwidth}  
    %     \centering 
    %     \includegraphics[width=\textwidth]{UHF_Pds.eps}
    %     \caption[]%
    %     {{\small UHF Band}}    
    %     \label{fig:QAM-64_Pds}
    % \end{subfigure}
%    \quad 
    \caption{Average Power Consumed v/s Target Rate with \textcolor{black}{rate forecaster. NYUSIM INH Indoor dataset} with LOS and NLOS. (a) and (b) are for dual bands (Sub-6 and mmWave), while (c) is for three bands (Sub-6, mmWave, and THz)}
\label{NYUSIM_Rateforecaster_LOSandNLOSIndoor_2and3bands}
\end{figure*}
To better understand power consumption with the channel forecaster, we plot the average power vs target data rate in Fig.
\ref{powerconsumed_vs_rate_ChanEst}. From  the graph, it is evident that with the channel based forecaster, the gap between average power of the LSTM and the greedy approach has reduced, and the current channel predictor performance is very close to that of the LSTM-based channel forecaster.

\textcolor{black}{We further validate the proposed approach on the NYUSIM outdoor data, as shown in Fig. \ref{nyusim_outdoor_lowmedhigh}, under the low, medium, and high blockage scenarios with rate forecaster.  The blockage scenarios are generated as discussed in Sec. IV-A-2. We can notice that the proposed LSTM based rate forecaster approach performs better than the Greedy, Current Channel prediction approach, and transformer based forecaster in terms of power consumption for all blockage scenarios. The reason for this is that the LSTM based approach has better forecasting ability as summarized in Table VII. We can also notice from the 2nd y-axis of Fig. \ref{nyusim_outdoor_lowmedhigh} (a)-(c) that the proportion of frames not meeting the target rate (say 2000 Mbps) is lowest for the low blockage scenario, and highest for the high blockage scenario. This is intuitive because the prediction worsens as the blockage increases. Nevertheless, on average, the LSTM based rate forecaster still performs better in terms of proportion of frames meeting the target rate. To quantify this, the distribution of rates actually achieved under the NYUSIM outdoor low and high blockage scenario is as shown in Fig. \ref{nyusim_cdf_low_highblockage_NYUSIMoutdoor}. We can notice that the distribution of rates below the threshold is approximately similar for all the approaches except that for the transformer based rate forecaster, especially in the high blockage scenario}.

\begin{figure*}[t!]
    \centering
    \captionsetup{format =hang}%
   % \vskip\baselineskip
    \begin{subfigure}[t]{0.33\textwidth}  
        \centering 
\includegraphics[width=\textwidth]{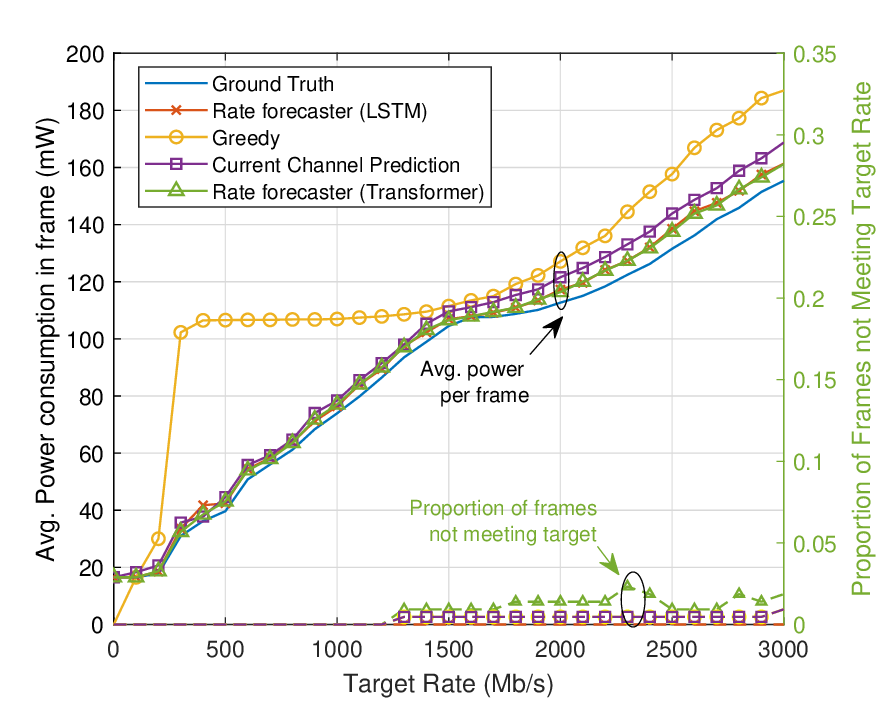}
        \caption[]%
        {{\small Rate forecaster}}    
        \label{fig:QPSK_Pds}
    \end{subfigure}
    \begin{subfigure}[t]{0.32\textwidth}  
        \centering 
        \includegraphics[width=\textwidth]{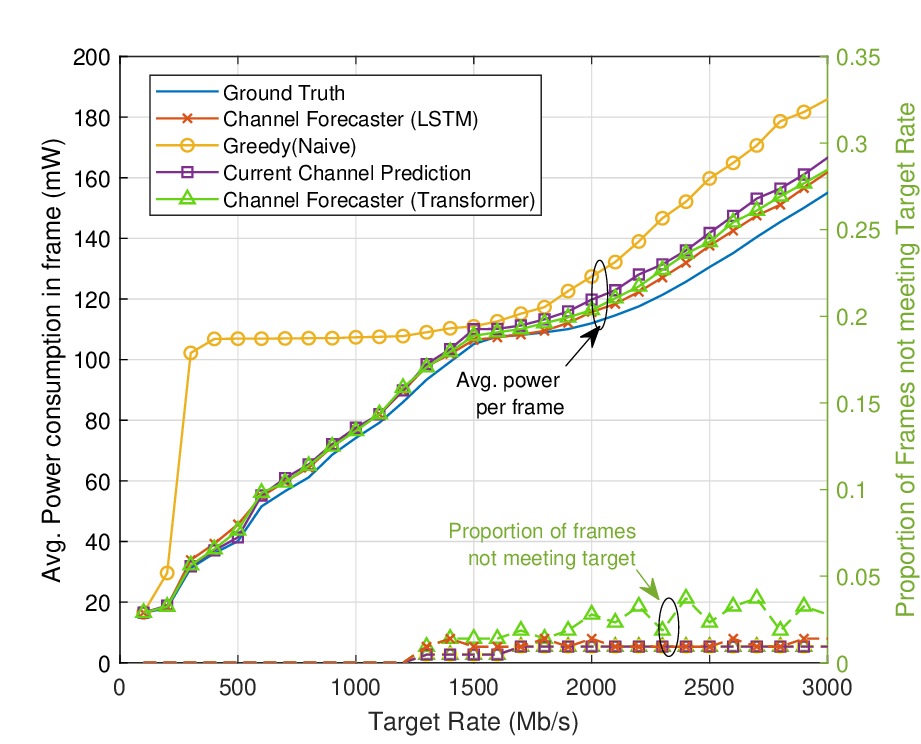}
        \caption[]%
        {{\small Channel forecaster }}    
        \label{fig:BPSK_Pds}
    \end{subfigure}
    \begin{subfigure}[t]{0.33\textwidth}  
        \centering 
        \includegraphics[width=\textwidth]{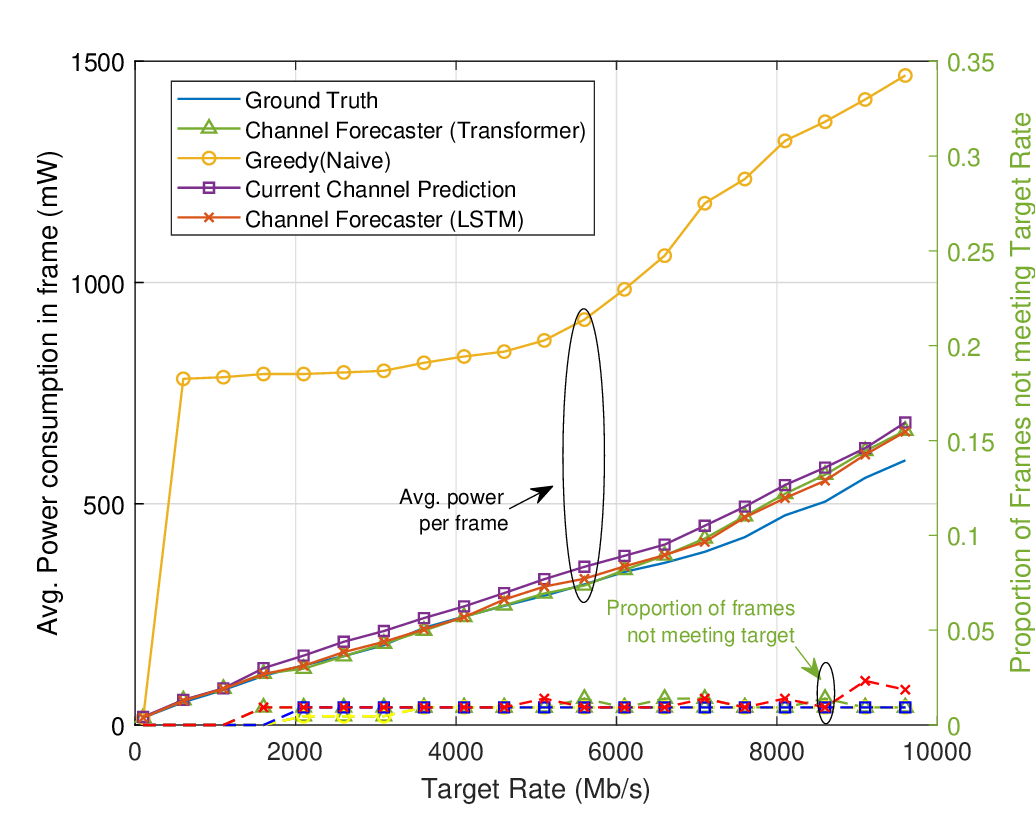}
        \caption[]%
        {{\small Channel forecaster (emulated sub-THz) }}    
        \label{fig:BPSK_Pds}
    \end{subfigure}
%    \hspace{0.1cm}
%    \quad
    %\hfill
%    \quad 
    \caption{Average Power Consumed v/s Target Rate with the multiband lab indoor dataset. Sub-figure (a) and (b) are for dual bands (Sub-6 and mmWave), while (c) is for three bands (Sub-6, mmWave, and emulated sub-THz)}
    \label{PconsPlot_labDataFastWalk}
\end{figure*}
\textcolor{black}{As mentioned earlier, in order to validate the proposed rate forecaster based energy aware band assignment approach, we test its performance on various indoor scenario datasets (described in Section IV - B, C). We would like to highlight that we consider only Sub-6 GHz and mmWave ISM bands for the indoor scenario (except in one scenario using NYUSIM as will be discussed later).  Fig. \ref{deepmimo_rateforecaster_LOSandNLOSIndoor}   shows the power consumption with the rate forecaster for the DeepMIMO Indoor I3 scenario  under LOS and NLOS scenario. We can notice that the proposed scheme closely follows the optimal approach in terms of power consumption, while the greedy approach has a significantly high power consumption even in indoor settings, followed by the current channel predictor. Since user mobility is low in indoor settings, the wireless channel parameters vary slowly and hence the LSTM and Transformer based approach closely follows the optimal approach, with a minor deviation for the transformer based approach in NLOS scenario in Fig. \ref{deepmimo_rateforecaster_LOSandNLOSIndoor} (b). This is also attributed to the fact that the forecasting errors using LSTM are relatively lower than those of the transformer as summarized in Table \ref{Table:summaryOfDeepMIMOIndoorPreds}.
%We can also notice that most of the frames achieve the target rate around 1000 Mbps except the current channel prediction scheme. This is mostly because it doesn't adapt its prediction based on the scenario. 
Furthermore, we observed that due to the high bandwidth available in the indoor ISM bands, the wide range of target rates was achieved mainly, and therefore we did not include the distribution plots of the rates for brevity.} 

\textcolor{black}{Fig. \ref{NYUSIM_Rateforecaster_LOSandNLOSIndoor_2and3bands}(a)-(c) shows the power consumption with the rate forecaster for the NYUSIM indoor InH scenario. 
We can notice that the proposed LSTM based rate forecaster approach is close to the optimal approach in terms of power consumption over the range of target rate, while the greedy approach and the current channel predictor have a significantly high power consumption. Moreover, the transformer based forecaster has small proportion of frames for which the target rate is not achieved. This observation is consistent for LOS and NLOS scenario in Fig. \ref{NYUSIM_Rateforecaster_LOSandNLOSIndoor_2and3bands}-(a) and (b), respectively. Given that the NYUSIM has a capability of generating the indoor channels at THz (142 GHz) frequency, we generate the three band channels i.e, Sub-6, mmWave, and THz with exactly the same user motion as in the case of dual bands but with the NLOS scenario. We can notice from Fig. \ref{NYUSIM_Rateforecaster_LOSandNLOSIndoor_2and3bands}-(c) that the proposed LSTM based rate forecaster still closely follows the optimal approach, on contrary, the greedy approach and current channel prediction has even higher gap in terms of power consumed per frame.} \textcolor{black}{We would like to highlight that similar performance in terms of power consumed per frame and proportion of frames not achieving target rate  was observed in the case of channel forecaster for the indoor DeepMIMO and NYUSIM scenarios, not included in the text for the sake of brevity.}

\textcolor{black}{Fig. \ref{PconsPlot_labDataFastWalk} describes the average power consumed per frame for different target rates with the experimentally collected multiband  indoor dataset. 
Fig. \ref{PconsPlot_labDataFastWalk}-(a) and (b) represent the rate forecaster and the channel forecaster with two bands, while Fig. \ref{PconsPlot_labDataFastWalk}-(c) represents the channel forecaster with three bands which includes the emulated sub-THz band.
As discussed in Section \ref{secIV}-C, for the testbed dataset collection, we moved the receiver cart back and forth multiple times at walking speed, to capture data close to real-world movements. Hence, this setting was a mix of LOS and NLOS i.e., LOS for portion of time when the receiving cart is in front of the transmitter and in NLOS for the rest of the time.  We can notice from  Fig. \ref{PconsPlot_labDataFastWalk} (a)-(b) that the average power consumed per frame with the proposed LSTM based forecaster follows closely to the optimal approach/ground truth for both the rate forecaster and channel forecaster schemes, respectively, while the greedy approach has considerable higher power consumption, \textcolor{black}{for instance $\sim$25 mW more at 2500 Mb/s, and significantly higher ($\sim$600 mW more) for three bands in Fig. \ref{PconsPlot_labDataFastWalk}-(c).} We would like to emphasize that the data collected using empirical testbed setup differs significantly from DeepMIMO-I3 and NYUSIM-InH settings. Specifically, the current setup employs an isotropic receiver antenna for mmWave band, which increases the likelihood of receiving multipath components compared to a directional antenna configuration. Moreover, given the relatively slow human walking speed in this experimental environment, we assume that the channel coherence time is considerably higher for mmWave \textcolor{black}{as compared to DeepMIMO and NYUSIM.}
%and even more so for sub-THz band.
%resulting in current channel predictor performance close to  proposed scheme and ground truth. 
Nevertheless, the proposed forecaster based techniques still outperform the current channel prediction approach by a few mW in dual band settings and even higher in a three band setup, \textcolor{black}{at the cost of $\sim$1\% of frames not meeting target rate}.
Moreover, although the transformer based forecaster closely follows our proposed approach, 
%it comes at the cost of  $\sim$3\% of frames not meeting target rate for rate forecaster (in Fig. \ref{PconsPlot_labDataFastWalk}-(a)), $\sim$5\% of frames not meeting target rate for channel forecaster (in Fig. \ref{PconsPlot_labDataFastWalk}-(b)), and $\sim$3\% of frames not meeting target rate for channel forecaster (in Fig. \ref{PconsPlot_labDataFastWalk}-(c)), all  in the range of 2000 - 3000 Mb/s target rates.
it comes at the cost of a fraction of frames failing to meet the target rate. Specifically, $\sim$3\% of frames fall below the target for the rate forecaster (Fig. \ref{PconsPlot_labDataFastWalk}-(a)), about 5\% for the channel forecaster (Fig. \ref{PconsPlot_labDataFastWalk}-(b)), and $\sim$3\% in Fig. \ref{PconsPlot_labDataFastWalk}-(c), all within the 2000–3000 Mb/s target rate range.}

\section{Conclusions}
%\vspace{-2mm}
\label{secVI}
\textcolor{black}{We propose a framework for energy aware channel assignments in multiband wireless communication systems where  high frequency bands can have significantly higher power consumption than low frequency bands \textcolor{black}{and are more prone to blockages}. In particular, we propose a framework where the average power consumption in a frame of $T$ slots in minimized while attempting to achieve a target average data rate in the frame. The channel assignments are done on the basis of forecasters which predict future channels/data rates based on recently learned history.}
We design rate forecaster(s) and channel forecaster(s) which forecasts $T$ direct multistep ahead using a stacked LSTM architecture.
%Moreover, we propose an iterative rate updating algorithm which updates the target rate based on current and future predicted rates in a frame. 
The proposed approach is validated on the publicly available  simulation based `DeepMIMO', and `NYUSIM' dataset for both outdoor and indoor settings, as well as in a multiband empirical testbed setup. %Moreover, we also propose a simple blockage segment generation algorithm.
\textcolor{black}{We propose a blockage segment generation algorithm such that the channel
realizations inherently capture the effects of blockage}
%We use O1-outdoor scenario of the publicly available  `DeepMIMO' dataset. 
We find that the rate forecaster  approach performs better than the channel forecaster. and LSTM based predictions  outperform the  Transformer-based  predictions  in NRMSE and NMAE. E.g., with a target rate of 1.5 Gb/s, we find that compared to the other approaches, the average power consumption per frame with the proposed approach is $\sim 100$ mW lower than a simple rate forecaster, which uses the current rate as the forecast for future rates, and $\sim 300$ mW lower than a greedy band assignment policy for the outdoor scenario and, up to 600 mW for indoor scenarios at 2.5 Gb/s target rate.  All of this is obtained  while  achieving the same distribution of rates below the threshold  as the optimal scheme. More generally, we find that power savings of 600 mW or more could possibly be obtained by similar band assignment approaches as the ability to forecast channels and rates improves over time, making the framework and analysis introduced in this paper helpful in improving the energy efficient operation of future wireless networks.

\section*{Acknowledgment}

\textcolor{black}{The authors thank Angel Rivas, an undergraduate student in the Department of Engineering at Boston College, for assistance with experimental data collection.}
\ifCLASSOPTIONcaptionsoff
  \newpage
\fi

\bibliographystyle{IEEEtran}
\bibliography{ref.bib} % References.bib file for BibTex

@ARTICLE{DL_mmwavebeam_blockage_prediction_usingsub6GHZ_TCOM2020,  author={Alrabeiah, Muhammad and Alkhateeb, Ahmed},  journal={IEEE Trans. Commun.},   title={Deep Learning for mmWave Beam and Blockage Prediction Using Sub-6 {GH}z Channels},   year={2020},  volume={68},  number={9},  pages={5504-5518},  doi={10.1109/TCOMM.2020.3003670},  ISSN={1558-0857},  month={Sep.},}

@ARTICLE{Beyond5G_MAC_HWN_ComMag_2018,  author={Cacciapuoti, Angela Sara and Sankhe, Kunal and Caleffi, Marcello and Chowdhury, Kaushik Roy},  journal={IEEE Commun. Mag.},   title={Beyond 5{G}: {TH}z-Based Medium Access Protocol for Mobile Heterogeneous Networks},   year={2018},  volume={56},  number={6},  pages={110-115},  doi={10.1109/MCOM.2018.1700924}}

@ARTICLE{THZ_last_piece_of_RF_OJCS,  author={Elayan, Hadeel and Amin, Osama and Shihada, Basem and Shubair, Raed M. and Alouini, Mohamed-Slim},  journal={IEEE Open J. Commun. Soc.},   title={Terahertz Band: The Last Piece of {RF} Spectrum Puzzle for Communication Systems},   year={2020},  volume={1},  number={},  pages={1-32},  doi={10.1109/OJCOMS.2019.2953633}}

@article{mmWave_and_THz_for6G_arxiv2021,
  title={Millimeter-wave and {T}erahertz Spectrum for 6{G} Wireless},
  author={Tripathi, Shuchi and Sabu, Nithin V and Gupta, Abhishek K and Dhillon, Harpreet S},
  journal={arXiv preprint arXiv:2102.10267},
  year={2021}
}

@ARTICLE{DL_based_predictive_band_switching_TWC2020,
  author={Mismar, Faris B. and Alammouri, Ahmad and Alkhateeb, Ahmed and Andrews, Jeffrey G. and Evans, Brian L.},
  journal={IEEE Trans. Wireless Commun.}, 
  title={Deep Learning Predictive Band Switching in Wireless Networks}, 
  year={2021},
  volume={20},
  number={1},
  pages={96-109},
  doi={10.1109/TWC.2020.3023397}}

@InProceedings{deepMIMo_ahmed_dataset_gen_paper,
author = {Alkhateeb, A.},
title = {{DeepMIMO}: A Generic Deep Learning Dataset for Millimeter Wave and Massive {MIMO} Applications},
booktitle = {Proc. of ITA workshop},
year = {2019},
pages = {1-8},
month = {Feb},
Address = {San Diego, CA}, }

@inproceedings{WiFi_assisted_60GHz_Mobicomm_2017,
author = {Sur, Sanjib and Pefkianakis, Ioannis and Zhang, Xinyu and Kim, Kyu-Han},
title = {WiFi-Assisted 60 {GH}z Wireless Networks},
year = {2017},
isbn = {9781450349161},
publisher = { },
address = {New York, USA},
url = {},
doi = {10.1145/3117811.3119870},
booktitle = {Proc. of the 23rd ACM MOBICOM},
pages = {28-40},
numpages = {3},
location = {Snowbird, Utah, USA},
}

@INPROCEEDINGS{2022_analysis_integrated_subb6_mmWave_Globecom2017,  author={Semiari, Omid and Saad, Walid and Bennis, Mehdi and Debbah, Merouane},  booktitle={GLOBECOM 2017 - 2017 IEEE Global Communications Conference},   title={Performance Analysis of Integrated Sub-6 {GH}z-Millimeter Wave Wireless Local Area Networks},   year={2017},  volume={},  number={},  pages={1-7},  doi={10.1109/GLOCOM.2017.8254610},  ISSN={},  month={Dec},}

@article{2020_max_system_throughput_Telkomika,
  title={Maximising system throughput in wireless powered sub-6 {GH}z and millimetre-wave 5{G} heterogeneous networks},
  author={Abdullah, Qazwan and Abdullah, Noorsaliza and Balfaqih, Mohammed and Shah, Nor Shahida Mohd and Anuar, Shipun and Almohammedi, Akram A and Salh, Adeeb and Farah, Nabil and Shepelev, Vladimir},
  journal={Telkomnika},
  volume={18},
  number={3},
  pages={1185--1194},
  year={2020}
}

@inproceedings{2017_asilomar_rate_Shroff,
  title={Rate-optimal power and bandwidth allocation in an integrated sub-6 {GH}z—Millimeter wave architecture},
  author={Hashemi, Morteza and Koksal, C Emre and Shroff, Ness B},
  booktitle={2017 51st Asilomar Conference on Signals, Systems, and Computers},
  pages={1207--1211},
  year={2017},
  organization={IEEE}
}

@ARTICLE{2022_bandswitch_molisch_TWC,
  author={Burghal, Daoud and Wang, Rui and Alghafis, Abdullah and Molisch, Andreas F.},
  journal={IEEE Trans.  Wireless Commun.}, 
  title={Supervised {ML} Solution for Band Assignment in Dual-Band Systems With Omnidirectional and Directional Antennas}, 
  year={2022},
  volume={21},
  number={9},
  pages={7550-7565},
  doi={10.1109/TWC.2022.3159762}}

@INPROCEEDINGS{2019_molish_duaband_shawdowing_correlation,  author={Burghal, Daoud and Nguyen, Sinh L. H. and Haneda, Katsuyuki and Molisch, Andreas F.},  booktitle={2019 IEEE Global Communications Conference (GLOBECOM)},   title={Dual Frequency Bands Shadowing Correlation Model in a Micro-Cellular Environment},   year={2019},  volume={},  number={},  pages={1-6},  doi={10.1109/GLOBECOM38437.2019.9013403},  ISSN={2576-6813},  month={Dec},}

@ARTICLE{2018_WCL_molish_rate_and_outage_analysis_dual_band,
  author={Burghal, Daoud and Molisch, Andreas F.},
  journal={IEEE Wireless Commun. Lett.}, 
  title={Rate and Outage Probability in Dual Band Systems With Prediction-Based Band Switching}, 
  year={2018},
  volume={7},
  number={5},
  pages={872-875},
  doi={10.1109/LWC.2018.2833860},
  ISSN={2162-2345},
  month={Oct},}

@INPROCEEDINGS{2020_6G_Summit_power_consumption_analysis_of_mmWave_SubTHz,  author={Skrimponis, Panagiotis and {et. al}},  booktitle={6G Wireless Summit},   title={Power Consumption Analysis for Mobile MmWave and Sub-{TH}z Receivers},   year={2020},  volume={},  number={},    doi={10.1109/6GSUMMIT49458.2020.9083793}}

@ARTICLE{Energy_constraint_sub6_goldsmith_TWC2005,  author={Shuguang Cui and Goldsmith, A.J. and Bahai, A.},  journal={IEEE Trans. Wireless Commun.},   title={Energy-constrained modulation optimization},   year={2005},  volume={4},  number={5},  pages={2349-2360},  doi={10.1109/TWC.2005.853882}}

@ARTICLE{system_level_Energy_sub6_TVLSI2007,  author={Li, Ye and Bakkaloglu, Bertan and Chakrabarti, Chaitali},  journal={IEEE Trans. Very Large Scale Integr. (VLSI) Syst.},   title={A System Level Energy Model and Energy-Quality Evaluation for Integrated Transceiver Front-Ends},   year={2007},  volume={15},  number={1},  pages={90-103},  doi={10.1109/TVLSI.2007.891095}}

@ARTICLE{M_zorzi_mmWave_TWC2017,  author={Abbas, Waqas Bin and Gomez-Cuba, Felipe and Zorzi, Michele},  journal={IEEE Trans. Wireless Commun.},   title={Millimeter Wave Receiver Efficiency: A Comprehensive Comparison of Beamforming Schemes With Low Resolution {ADC}s},   year={2017},  volume={16},  number={12},  pages={8131-8146},  doi={10.1109/TWC.2017.2757919}}

@misc{keras,
	author = {Chollet, Franc¸ois and others},	
	title = {{Keras}},
	howpublished = {Available: {https://github.com/fchollet/keras}},
	note = {},
	year = 2015
}

@ARTICLE{integrated_mmWave_sub6_roadmap_MDebbah_Comm_magazine2019,  author={Semiari, Omid and Saad, Walid and Bennis, Mehdi and Debbah, Merouane},  journal={IEEE Wireless Commun.},   title={Integrated Millimeter Wave and Sub-6 {GH}z Wireless Networks: A Roadmap for Joint Mobile Broadband and Ultra-Reliable Low-Latency Communications},   year={2019},  volume={26},  number={2},  pages={109-115},  doi={10.1109/MWC.2019.1800039},  ISSN={1558-0687},  month={Apr.},}

@INPROCEEDINGS{Bandswitch_DRL_UAV_hamadi_VTC2021,  author={Fontanesi, Gianluca and Zhu, Anding and Ahmadi, Hamed},  booktitle={Proc. of IEEE 94th VTC-Fall},   title={Deep Reinforcement Learning for Dynamic Band Switch in Cellular-Connected  {UAV}},   year={2021},  volume={},  number={},  doi={10.1109/VTC2021-Fall52928.2021.9625502},  ISSN={2577-2465},  month={Sep.},}

@techreport{3gpp_36_331,
 author = {3GPP},
 day = {},
 institution = {{3rd Generation Partnership Project (3GPP)}},
 month = {Apr.},
 note = {Version 15.5.1},
 number = {36.331},
 title = {{Evolved Universal Terrestrial Radio Access (E-UTRA); Radio Resource Control (RRC); Protocol specification}},
 type = {TS},
 url = {},
 year = {2019}
}

@INPROCEEDINGS{Our_ccnc_paper2023,
  author={Soni, Brijesh and Govindasamy, Siddhartan and Patel, Dhaval K.},
  booktitle={Proc. of 20th IEEE CCNC}, 
  title={Deep Learning aided Energy Efficient Band Assignment in Multiband Heterogeneous Networks}, 
  year={2023},
  volume={},
  number={},
  pages={690-691},
  doi={10.1109/CCNC51644.2023.10060279}}

@ARTICLE{Miguel_TWC2023,
  author={Adamu, Paul Ushiki and López-Benítez, Miguel and Zhang, Jiayi},
  journal={IEEE Trans. Wireless Commun.}, 
  title={Hybrid Transmission Scheme for Improving Link Reliability in mm{W}ave {URLLC} Communications}, 
  year={2023},
  volume={},
  number={},
  doi={10.1109/TWC.2023.3241792}}

@article{transforemr_for_time_series_AAAI23,
  title={Are Transformers Effective for Time Series Forecasting?},
  author={Ailing Zeng and Muxi Chen and Lei Zhang and Qiang Xu},
  journal={Proceedings of the AAAI Conference on Artificial Intelligence},
  year={2023}
}

@article{vaswani2017attention,
  title={Attention is all you need},
  author={Vaswani, Ashish and et. al},
  journal={Advances in {N}eural {I}nformation {P}rocessing Systems},
  volume={30},
  year={2017}
}

@ARTICLE{2024_adaptive_RIS_blockage_mmWave,
  author={Zhang, Bibo and Filippini, Ilario and Lian, Zhuxian and Su, Yinjie},
  journal={IEEE Commun. Lett.}, 
  title={Adaptive Obstacle-Aware {RIS} Switch for Mobile mm{W}ave Access Networks}, 
  year={2024},
  volume={28},
  number={5},
  pages={1216-1220},
  doi={10.1109/LCOMM.2024.3376713}}

@INPROCEEDINGS{2020_Globecom_sub6_mmWave_blockage,
  author={Göttsch, Fabian and Kaneko, Megumi},
  booktitle={IEEE Global Communications Conference}, 
  title={Deep Learning-based Beamforming and Blockage Prediction for Sub-6 {GH}z/mm{W}ave Mobile Networks}, 
  year={2020},
  volume={},
  number={},
  pages={1-6},
  doi={10.1109/GLOBECOM42002.2020.9322404}}

@ARTICLE{2023_CL_LOS_NLOS_identification_Sub6,
  author={Zhu, Yasong and Xu, Bing and Wang, Jiabao and Li, Yuqing and Qi, Wangdong},
  journal={IEEE Commun. Lett.}, 
  title={A Simple Efficient Lightweight CNN Method for {LOS/NLOS} Identification in Wireless Communication Systems}, 
  year={2023},
  volume={27},
  number={6},
  pages={1515-1519},
  doi={10.1109/LCOMM.2023.3265272}}

@ARTICLE{2020_TWC_LOS_NLOS_identification_Sub6,
  author={Huang, Chen and et. al},
  journal={IEEE Trans. Wireless Commun.}, 
  title={Machine Learning-Enabled {LOS/NLOS} Identification for {MIMO} Systems in Dynamic Environments}, 
  year={2020},
  volume={19},
  number={6},
  pages={3643-3657},
  doi={10.1109/TWC.2020.2967726}}

@ARTICLE{2021_WCL_mmWave_Thz_linkPrediction_analytical,
  author={Kalør, Anders E. and Simeone, Osvaldo and Popovski, Petar},
  journal={IEEE Wireless Commun. Letter}, 
  title={Prediction of mm{W}ave/{TH}z Link Blockages Through Meta-Learning and Recurrent Neural Networks}, 
  year={2021},
  volume={10},
  number={12},
  pages={2815-2819},
  doi={10.1109/LWC.2021.3118269}}

@misc{nyusim4,
	author = {},	
	title = {{NYUSIM Ver4.0}},
	howpublished = {Accessed: Nov. 2023. [Online]. Available: {\href{https://wireless.engineering.nyu.edu/nyusim/}{https://wireless.engineering.nyu.edu/nyusim}}},
	note = {},
	year = {}
}

@INPROCEEDINGS{our_globecom23_paper,
  author={Soni, Brijesh and Govindasamy, Siddhartan and Patel, Dhaval K.},
  booktitle={Proc. of IEEE Global Communications Conference (GLOBECOM)}, 
  title={Rate Forecaster based Energy Aware Band Assignment in Multiband Networks}, 
  year={2023},
  volume={},
  number={},
  pages={6723-6728},
  doi={10.1109/GLOBECOM54140.2023.10437500}}

@ARTICLE{2024TCOM_multiband_survey,
  author={Aboagye, Sylvester and Amin Saeidi, Mohammad and Tabassum, Hina and Tayyar, Yamin and Hossain, Ekram and Yang, Hong-Chuan and Alouini, Mohamed-Slim},
  journal={IEEE Trans Commun.}, 
  title={Multi-Band Wireless Communication Networks: Fundamentals, Challenges, and Resource Allocation}, 
  year={2024},
  volume={72},
  number={7},
  pages={4333-4383},
  doi={10.1109/TCOMM.2024.3366816}}

@ARTICLE{2023_comm_magazine_multiband,
  author={Saeidi, Mohammad Amin and Tabassum, Hina and Alouini, Mohamed-Slim},
  journal={ IEEE Commun. Mag. }, 
  title={Multi-Band Wireless Networks: Architectures, Challenges, and Comparative Analysis}, 
  year={2024},
  volume={62},
  number={1},
  pages={80-86},
  doi={10.1109/MCOM.003.2300078}}

@inproceedings{saha2017detailed,
  title={A detailed look into power consumption of commodity 60 {GH}z devices},
  author={Saha, Swetank Kumar and Siddiqui, Tariq and Koutsonikolas, Dimitrios and Loch, Adrian and Widmer, Joerg and Sridhar, Ramalingam},
  booktitle={Proc. of IEEE 18th International Symposium on A World of Wireless, Mobile and Multimedia Networks (WoWMoM)},
  year={2017},
}

\vfill

\end{document}